\documentclass[
  reprint,
  aps,
  prx,
  amsmath,amssymb
]{revtex4-2}

\usepackage{braket}
\usepackage{bm}

\usepackage{graphicx}
\usepackage{dcolumn} 
\usepackage{float}

\usepackage{orcidlink}
\usepackage{xcolor}
\usepackage{soul}
\usepackage{comment}
\usepackage{hyperref}

\newcommand{\csz}[1]{\textcolor{orange}{[SZ: #1]}}
\newcommand{\cstsz}[1]{\textcolor{orange}{[SZ: \st{#1}]}}

\newcommand{\mhl}[1]{\textcolor{black}{#1}}
\newcommand{\rhl}[1]{\textcolor{black}{#1}}
\begin{document}

\title{Transduction-Enabled Superconducting Quantum Repeater: Toward Deterministic Entanglement Distribution with High-Fidelity Gates}

\author{Francesco Fiorini$^{1,2}$\orcidlink{0000-0002-5572-3623}}
\author{Jing Wu$^{2}$\orcidlink{0000-0002-4946-0732}}
\author{Andrew Cameron$^{2}$\orcidlink{0000-0002-3739-9313}}
\author{Changqing Wang$^{2}$\orcidlink{0000-0001-9807-3045}}
\author{Doga M. Kurkcuoglu$^{2}$\orcidlink{0000-0003-1109-7074}}
\author{Rosario G. Garroppo$^{1}$\orcidlink{0000-0001-7465-6019}}
\author{Michele Pagano$^{1}$\orcidlink{0000-0003-1706-4994}}
\author{Silvia Zorzetti$^{2, 3}$\orcidlink{0000-0002-3208-3387}}
\email{zorzetti@ieee.org}

\affiliation{$^{1}$Department of Information Engineering, University of Pisa, Pisa, Italy}
\affiliation{$^{2}$Fermi National Accelerator Laboratory, Batavia, IL, USA}
\affiliation{$^{3}$Department of Physics and Astronomy, Northwestern University, Evanston, IL, USA}

\date{\today}

\begin{abstract}

Long-distance entanglement distribution is hindered by photon loss in optical fibers and the no-cloning theorem. Optical quantum repeater (QR) protocols rely on Bell state measurements (BSMs), they are intrinsically limited to probabilistic photon operations and fail 50\% of the time. We propose a hybrid approach to building quantum repeaters that combines the high transmission speed of photonic qubits in optical fiber with the high-fidelity quantum processing capabilities enabled by superconducting circuits. The transduction-enabled superconducting QR (TESQR) architecture eliminates the need for probabilistic BSMs and
allows deterministic processing operations. The TESQR framework always yields a final state at the remote nodes rather than aborting on photon loss\mhl{, manifesting deterministic entanglement distribution within certain parameter regimes.} We evaluate the performance by assessing output-state fidelities and success probabilities of entanglement distribution using realistic noise models.  Additionally, we integrate an entanglement purification scheme and evaluate the performance through numerical simulations in QuTiP environment. \mhl{Our results show that, for entanglement swapping, the proposed scheme improves the entanglement distribution rate by an average of 63\% and by up to 159\% compared with photonic-only architectures. Moreover, after purification, the end-to-end fidelities exceed 0.8 over distances up to 20~km.}

\end{abstract}

\maketitle

\section{Introduction}
\label{sec:intro}
Entanglement distribution is a fundamental component of future quantum internet networks. Realizing long-distance quantum communication remains a major challenge due to the exponential attenuation of photons in optical fibers and operational errors in intermediate processing nodes. Moreover, qubits cannot be amplified or cloned like classical signals without destroying their quantum information, in accordance with the no-cloning theorem. For these reasons, quantum repeaters (QRs)~\cite{QRsurvey} are needed, as they split the end-to-end channel into multiple shorter segments and employ various error-suppression techniques to securely extend the communication range.

Over the past two decades, numerous QR architectures have been proposed. The Duan-Lukin-Cirac- Zoller (DLCZ) protocol~\cite{dlcz} employs atomic-ensemble memories and heralded Raman emissions, together with optical Bell-state measurements (BSMs) on the \emph{readout} photons for entanglement swapping. Although it paved the way for the first experimental demonstrations, the DLCZ protocol exhibits a low per-trial success probability (typically spanning the $10^{-6}$-- $10^{-4}$ range) and requires phase stabilization and high detection efficiencies, limiting the achievable entanglement distribution rate~\cite{atomicensemble}.

Other protocols adopt purely photonic strategies, avoiding stationary memories and instead relying on large photonic cluster states with adaptive single-qubit measurements~\cite{QRsurvey}. Examples include multiplexed photonic cluster approaches~\cite{munro_quantum_2010} and all-photonic repeater schemes based on CSS-code trees and homodyne detection~\cite{azuma_all-photonic_2015}. While these approaches eliminate the need for matter qubit storage, they do so at the cost of very large photon overheads and high sensitivity to channel and component losses~\cite{opticalQRchallenge}.

Hybrid approaches attempt to combine the advantages of both matter and photonic systems. Van Loock~\emph{et al.}~\cite{Loock} proposed using dispersive interactions between matter qubits (e.g., atoms in cavities) and bright coherent pulses, where a single optical pulse sequentially traverses two cavities and acquires phase rotations conditional on the state of each qubit. A homodyne measurement on the outgoing field then heralds entanglement, with success probabilities exceeding those of purely photonic schemes. Subsequent work showed that, while such schemes enable faster entanglement generation, the initial fidelity is sensitive to channel loss and cavity decay, typically requiring many purification rounds that increase system complexity and latency~\cite{ladd_hybrid_2006,rateanalysis}.
Trapped-ion-based nodes provide another hybrid route: photons entangled with the ions’ internal states are interfered to herald remote entanglement, while deterministic local gates (e.g., Mølmer–Sørensen) enable high-fidelity swapping and readout. Variants employing co-trapped species address wavelength matching to telecom bands and have demonstrated promising simulated ranges~\cite{QRtrappedions,santra_quantum_2019}.

Despite these advances, a recurring bottleneck across most implementations is the optical Bell state measurement (BSM) used for link-level entanglement generation or swapping. Its success probability in linear-optical implementations is fundamentally capped at 50\%, since only two of the four Bell states can be distinguished~\cite{deterministicswapping,calsamiglia_maximum_2001}. In practice, the entanglement rates are further reduced by photonic losses and by the need for precise synchronization to enable optical interference~\cite{teleportation,opticalQRchallenge}. 
Alternative strategies do exist to exceed the 50\% linear-optical bound, such as those based on hyperentanglement, nonlinear interactions, or ancillary resources~\cite{BSMmorethan50,PRXQuantum.4.040322}. However, these solutions introduce additional experimental overheads and trade-offs. 

Recent years have witnessed significant progress in bidirectional microwave-optical quantum transduction, realized on several platforms such as electro-optic crystalline cavities~\cite{wang_high-efficiency_2022,Jing2}, electro/piezo-optomechanics~\cite{Zhao2025}, rare-earth ions~\cite{Xie2025}, atomic assembly, and others~\cite{lauk2020perspectives}. They have provided a useful platform to combine the high-fidelity quantum operations of superconducting circuits with the room-temperature communication capabilities via optical flying photons.
Here, we propose a transduction-enabled superconducting quantum repeater (TESQR), a novel hybrid architecture that extends the outlook of previous works~\cite{QRFiorini} and draws inspiration from the general setup described in~\cite{DiVincenzo2011Patent}. This scheme differs from existing solutions by 
eliminating probabilistic optical BSM during both entanglement generation and swapping, thereby enabling 
deterministic processing operations. With TESQR, we combine the advantages of superconducting quantum computing devices and optical fibers, i.e., integrating fast gate operations with low-loss communication. It also overcomes the intrinsic limitation of 50\% Bell state discrimination and synchronization challenges, while improving the final output fidelity. 

In this work, we analyze the processing performance of the hybrid architecture by evaluating: (i) the output fidelity, and (ii) the protocol's success probability for entanglement distribution.
This analysis is conducted by evaluating various noise models. To this end, we characterize the processing operations using several figures of merit for the different setup components. Since our primary objective is to assess the feasibility and potential advantages of the proposed hybrid scheme at the level of quantum processing operations, we do not model imperfections in quantum memories and instead assume ideal memory performance, which is consistent with previous studies 
~\cite{entangling,rateanalysis}. Specifically, we assume that all operations are completed within a time $t < t_{\mathrm{coh}}$, where $t_{\mathrm{coh}}$ denotes the memory coherence time. \mhl{Furthermore, we highlight that our analysis focuses on quantum-processing performance metrics rather than on resource costs. A systematic resource-cost investigation would require defining non-standard, cross-platform metrics to account for the fundamentally different physical constituents involved (superconducting qubits, transducers, photodetectors, etc.), and is therefore left for dedicated future work.}

Finally, we propose 
entanglement purification in 
a superconducting quantum processing unit (QPU) before the swapping step. 
For this purpose, we develop a QuTiP-based simulation framework tailored to the considered network scenario. Numerical results demonstrate the effectiveness of the proposed architecture, achieving end-to-end fidelity values exceeding 0.8 for link distances up to 20~km between the two parties under realistic processing imperfections.

To summarize, the main contributions of this work are as follows:

\begin{itemize}
    \item We propose a novel hybrid quantum repeater architecture (TESQR) that leverages quantum transduction \mhl{to enable deterministic entanglement distribution within certain parameter regimes}.
    
    \item We develop a comprehensive analytical model of the network subsystems, including a detailed noise characterization, and evaluate the TESQR performance in terms of fidelity and success probability.
    
    \item We introduce a performance enhancement strategy based on an additional dedicated purification stage and validate the resulting improvement through a publicly available software implementation~\cite{simulatorTESQR}.
\end{itemize}

The paper is structured as follows. In Section~\ref{sec:TESQRArchitecture}, we describe the TESQR architecture; in Section~\ref{sec:model}, we present the analytical model of the various noisy operations in the network scenario. Section~\ref{sec:performanceval} evaluates the performance of the TESQR architecture for entanglement swapping in terms of output fidelity and success probability. In Section~\ref{sec:simresults}, we introduce the purification scheme and report the corresponding numerical results obtained with the QuTiP simulation framework. Finally, Section~\ref{sec:conclusion} concludes the manuscript and outlines future development directions.

\section{TESQR Architecture}
\label{sec:TESQRArchitecture}

\begin{figure}[htb]
\centering
\includegraphics[width=\linewidth]{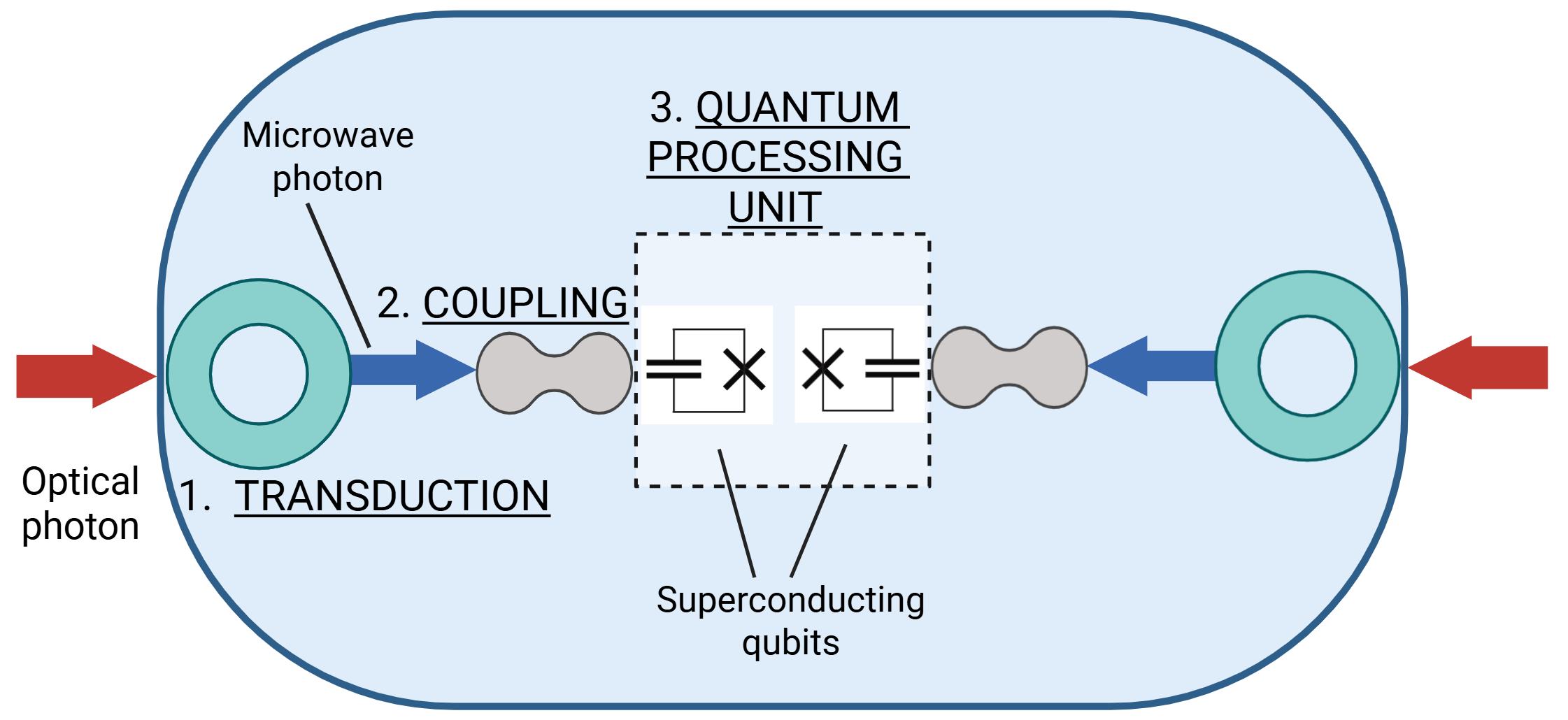}
\caption{Illustration of TESQR node architecture performing joint measurement on two optical photons via transduction, coupling, and superconducting QPU.}
\label{fig:QRarchitecture}
\end{figure}

The protocol considers entangled pairs between a photonic qubit (Fock-state, single-rail encoding) and superconducting circuits. The TESQR node is built on a functional layer comprising the three main components, as illustrated in Fig.~\ref{fig:QRarchitecture}:
\begin{enumerate}
    \item \textbf{Transduction--} After propagation via optical fibers, the photons are converted into microwave photons via a quantum transducer.
    \item \textbf{Coupling--} After the conversion, the microwave photon couples directly to a superconducting QPU.
    \item \textbf{Swapping--} Entanglement swapping is performed locally on the QPU, using microwave gates followed by measurements.
\end{enumerate}


Key advantages of this approach are:
\begin{itemize}

\item Leveraging the high-fidelity gate operations achievable with superconducting circuits (single-qubit error rates below $10^{-4}$)~\cite{li_error_2023,CNOT,engineeringguide,superconducting_2020};
\item Combining the benefits of using low-loss optical photons at standard telecom frequencies for transmission with the robust microwave-domain processing capabilities of stationary superconducting qubits, made possible by recent advances in quantum transduction and signal processing~\cite{Jing};
\item Overcoming the intrinsic 50\% theoretical limitation of linear-optical BSMs~\cite{entangpurification} by employing a deterministic Bell-measurement scheme for the entanglement distribution at the repeater node.
\item Guaranteeing a final end-to-end entangled state, which could be noisy and suitable for further processing, such as purification. Heralded photonic schemes, instead, inherently discard failures and thus prevent performance improvement \cite{entangling}.
\end{itemize}

Table~\ref{tab:tesqr_stages} provides a summary of the main sequential steps required to implement entanglement swapping within the network scenario under consideration.

\begin{table*}[htb]
\centering
\caption{Network and TESQR node stages and corresponding processes to perform entanglement swapping.}
\label{tab:tesqr_stages}
\renewcommand{\arraystretch}{1.7}
\begin{tabular}{c c c}
\hline\hline
Stage & Process & Description \\[4pt]
\hline
Communication -- Stage 1 & Source & \parbox[t]{10cm}{Alice and Bob locally generate entangled Bell-like states.} \\
Communication -- Stage 2 & Transmission & \parbox[t]{10cm}{The photonic qubits propagate through an optical fiber of length $l$, characterized by an attenuation length $l_{\mathrm{att}}$.} \\
TESQR -- Stage 1 & Transduction & \parbox[t]{10cm}{Optical-to-microwave frequency conversion characterized by thermal noise with mean photon occupancy $\bar{n}$ and transduction efficiency $\eta_t$.} \\
TESQR -- Stage 2 & Coupling & \parbox[t]{10cm}{The converted microwave photons couple to the superconducting qubits.} \\
TESQR -- Stage 3 & Swapping & \parbox[t]{10cm}{Local CNOT and Hadamard gates and measurements are applied to the superconducting qubits to swap the entanglement between Alice and Bob.} \\[0.7cm]
\hline\hline
\end{tabular}
\end{table*}

\section{Model Analysis}
\label{sec:model}
The network model under consideration involves the two communication parties, Alice and Bob, each equipped with entanglement sources, aiming to perform entanglement distribution. The TESQR architecture is the intermediate repeater node located halfway along the optical fiber connecting Alice and Bob. It executes entanglement swapping and entanglement purification. 

\begin{figure*}[htb]
    \centering
    \includegraphics[width=0.7\linewidth]{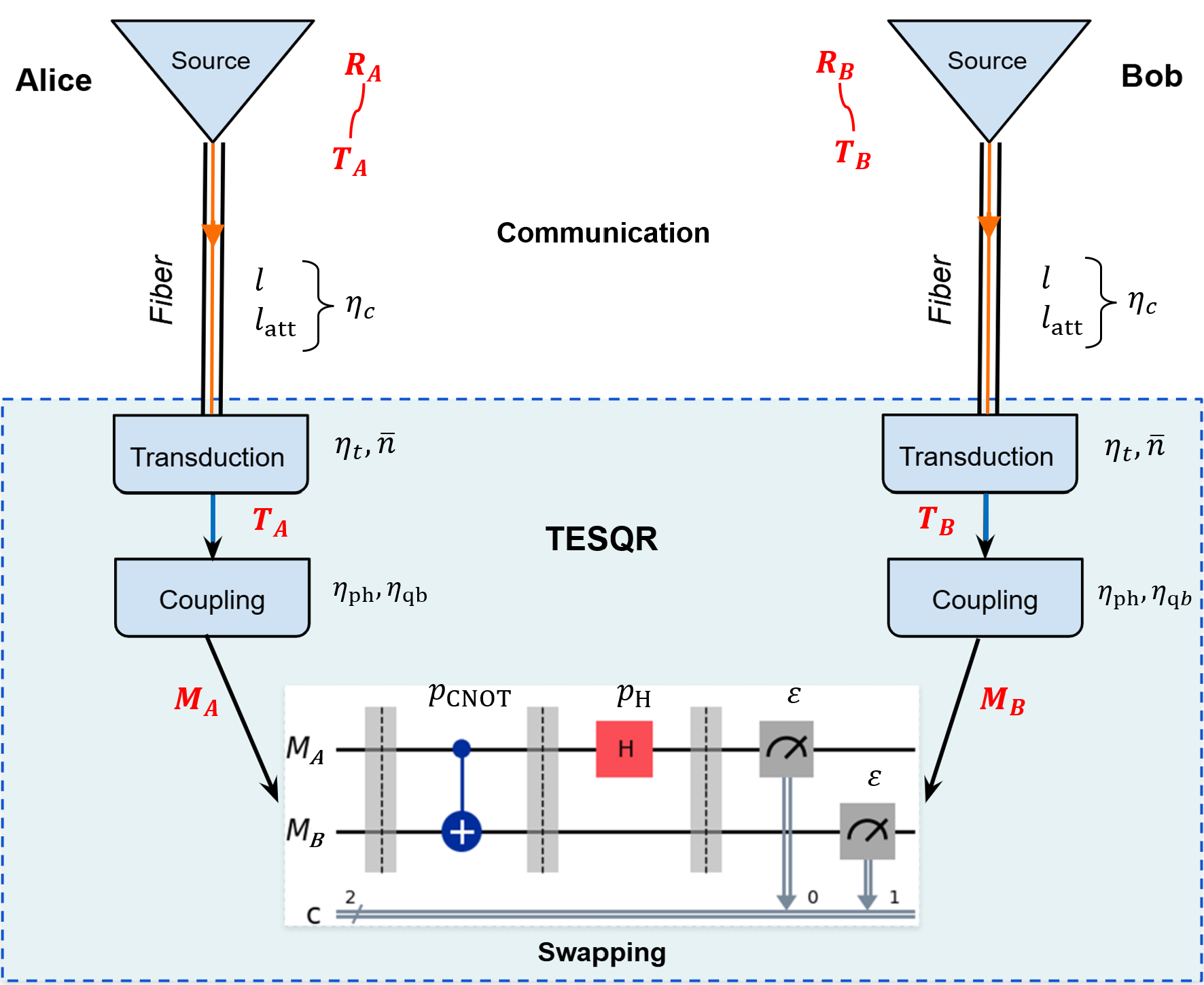}
    \caption{Schematic of the network architecture. The setup features Alice and Bob at the endpoints, each producing an entangled pair: a transmitted photonic qubit ($T_A$ or $T_B$) in single-rail Fock-state encoding and a remote superconducting qubit ($R_A$ or $R_B$). Photonic qubits travel through optical fibers of length $l$ with transmissivity $\eta_\mathrm{c} = e^{-l/l_\text{att}}$ ($l_\text{att} = 22$ km, equivalent to 0.2 dB/km at 1550 nm). At the TESQR node, optical-to-microwave conversion occurs via electro-optic transduction with efficiency $\eta_t$ and thermal noise photon number $\bar{n}$. Microwave photons couple to superconducting qubits $M_A$ and $M_B$ with photon availability $\eta_\text{ph}$ and transition efficiency $\eta_\text{qb}$. Gate-based entanglement swapping uses a CNOT gate (error $p_\text{CNOT}$), Hadamard gate (error $p_H$), and POVM measurements (error $\varepsilon$). \mhl{Optical and microwave photon paths are highlighted in orange and blue, respectively.}}
    \label{fig:networksetup}
\end{figure*}

Here, we analytically examine the effect of each processing step on the source states, modeling the various operations in a cascade with appropriate noise models. Figure~\ref{fig:networksetup} illustrates the network architecture under consideration. The model parameters are discussed in next subsections and summarized in~Table \ref{tab:networkparameters}.

\begin{table*}[htb]
\centering
\caption{Network model parameters definition.}
\renewcommand{\arraystretch}{1.2}
\begin{ruledtabular}
\begin{tabular}{c c c}
Process & Parameter & Definition \\ 
\hline
Transmission & $l$ & Fiber length (km) between Alice/Bob--TESQR \\
Transmission & $l_{\mathrm{att}}$ & Fiber attenuation length (km) \\
Transmission & $\eta_\mathrm{c}$ & Fiber transmissivity \\
Transduction & $\eta_\mathrm{t}$ & Transduction efficiency \\
Transduction & $\bar{n}$ & Mean noise photon number in the transduction process \\
Transduction & $N_\mathrm{t}$ & Transduction output added noise \\
 Transduction & $T_{\mathrm{sys}}$ & Transduction system temperature (K)  \\
 Transduction & $\zeta_{\mathrm{m}}$ &  Microwave extraction efficiency \\
 Transduction & $\zeta_{\mathrm{o}}$ &  Optical extraction efficiency\\
 Coupling & $\eta_{\mathrm{ph}}$ &  Probability of microwave photon availability\\
Coupling & $\eta_{\mathrm{qb}}$ & Probability of superconducting qubit excitation upon photon absorption  \\
Swapping & $p_{\mathrm{H}}$ & Hadamard gate error rate \\
Swapping & $p_{\mathrm{CNOT}}$ & CNOT gate error rate \\
Swapping & $\varepsilon$ & POVM measurement error rate\\
\end{tabular}
\end{ruledtabular}
\label{tab:networkparameters}
\end{table*}

The network scenario depicted in Fig. \ref{fig:networksetup} highlights the symmetry of the setup in the two segments Alice–TESQR and Bob–TESQR. Accordingly, to avoid overburdening the mathematical notation, in the following we consider only one of the two equivalent links and omit, where possible, the numerical subscripts. \mhl{A numerical treatment of the asymmetric scenario, in which the parameters of the two segments are independently varied, is provided and discussed in Appendix \ref{app:asymmetric}.}

\subsection{Source States}
\label{sec:sourcestate}
The sources generate Bell-type entangled states composed of transmitted photonic subsystems ($T_A$ for Alice and $T_B$ for Bob) and a remote superconducting subsystem ($R_A$ for Alice and $R_B$ for Bob). For subsystem $T$, we adopt a single-rail Fock state encoding ($\ket{0}, \ket{1}$), defined by the absence ($\ket{0}$) or presence ($\ket{1}$) of a single photon in a bosonic mode~\cite{entangling}. This choice of photonic encoding is motivated by the subsequent physical feasibility of the coupling process, which would otherwise be hindered by alternative encoding methods (such as time-bin encoding)~\cite{FioriniICC2026Fidelity}. 
Moreover, in standard implementations with photonic BSM, this encoding requires single-photon interference at the repeater node, thereby allowing a higher entangling probability compared to double-photon interference~\cite{entangling,krutyanskiy_light-matter_2019,yu_entanglement_2020}. 

For the remote stationary qubit subsystem $R$, let $\ket{g}$ and $\ket{e}$ denote the ground and excited states, respectively. Accordingly, the source state (for Alice or Bob, equivalently) is defined as:
\begin{equation}
\label{eq:source}
|\psi\rangle_{TR} = \frac{1}{\sqrt{2}}\Bigl(|0\rangle_T\otimes|g\rangle_R + |1\rangle_T\otimes|e\rangle_R\Bigr).
\end{equation}
The generation of entangled states of the form in Eq.~(\ref{eq:source}) is experimentally achievable, as documented in~\cite{source1}.  
\mhl{Imperfections are introduced from the quantum channel onwards, while the modeling of imperfect sources is intentionally left outside the scope of this work. This choice reflects the aim of the paper, which is to evaluate the performance of the proposed hybrid repeater node independently of the specific physical implementation of the other protocol components.}

To simplify the derivation below, it is convenient to introduce the displacement operator acting in the photonic mode, as in~\cite{olivares_introduction_2021}:
\(
D_T(\xi)=\exp\Bigl(\xi a^\dagger - \xi^* a\Bigr),
\)
where, \( a^\dagger \) and \( a \) are the creation and annihilation operators, and \( \xi \) is a complex phase-space variable. Its matrix elements are defined as
\(
X_{mn}(\xi)=\langle m|D_T(\xi)|n\rangle,\quad m,n\in\{0,1\}.
\)

Similarly to the quantum characteristic function, we now define the reduced characteristic operator by applying the displacement operator to the subsystem \(T\) and then tracing the photonic mode:
\begin{equation}
\rho_R^{\mathrm{source}}(\xi)=\operatorname{Tr}_T\Bigl\{\rho_{TR}^{\mathrm{source}}\,D_T(\xi)\Bigr\},
\end{equation}
where $\rho_{TR}^{\mathrm{source}} = |\psi\rangle_{TR}\langle\psi|_{TR}$. We may also recover the state from the reduced operator by the Glauber equation~\cite{olivares_introduction_2021}:
\begin{equation}
    \rho_{TR}^{\text{source}} = \frac{1}{\pi} \int \rho_R^{\text{source}}(\xi) D_T^\dagger(\xi) \, d^2\xi.
    \end{equation}

\subsection{Fiber Transmission}
\label{sec:fiber}
The optical fiber channel between the source and TESQR, of length $l$, is modeled as a Gaussian pure-loss bosonic channel~\cite{channels,QRsurvey}, with associated transmissivity $\eta_C \in [0,1]$, which physically corresponds to the probability of photon survival. For simplicity, we include other sources of attenuation (e.g., connector losses) in the channel transmissivity. This parameter is related to the fiber attenuation length $l_\mathrm{att}$ through the exponential relation:
\begin{equation}
\eta_\mathrm{c}=e^{-l/l_\mathrm{att}}.
\end{equation}  
We assume a standard telecommunication wavelength of 1550~nm, with $l_\mathrm{att}=22$~km, corresponding to an attenuation coefficient of 0.2~dB/km. The attenuation coefficient chosen for this analysis is aligned with realistic standard for fiber-based deployment~\cite{QRsurvey}.

The Gaussian pure-loss channel with transmissivity $\eta_\mathrm{c}$ transforms the displacement operator as
\(
D_T(\xi) \longrightarrow D_T\Bigl(\sqrt{\eta_\mathrm{c}}\,\xi\Bigr)e^{-\frac{1-\eta_\mathrm{c}}{2}|\xi|^2}\,.
\)
The state after transmission through the fiber is expressed as:
\begin{equation}
\rho_R^{\mathrm{ch}}(\xi) = \rho_R^{\mathrm{source}}\Bigl(\sqrt{\eta_\mathrm{c}}\,\xi\Bigr)e^{-\frac{1-\eta_\mathrm{c}}{2}|\xi|^2}\,.
\end{equation}
The explicit analytical expressions, obtained both here and in the subsequent steps, are reported in Appendix~\ref{app:densitymatrices}.

\subsection{Transduction}
\label{sec:transduction}
For the optical-microwave quantum transduction process, we refer to the electro-optic implementation approach~\cite{wang_high-efficiency_2022,Jing2}. This process can be modeled as a single-mode Gaussian quantum channel with transmissivity \( \eta_\mathrm{t} \in [0, 1] \) and 
added noise:  
\begin{equation} N_\mathrm{t} = (1 - \eta_\mathrm{t}) (\bar{n} + \tfrac{1}{2}), \end{equation}  
where \( \bar{n} \) denotes the mean noise photon number in the beamsplitter model~\cite{Jing2}. The latter also determines the efficiency-noise tradeoff of the transducer and depends on physical parameters such as system temperature ($T_{\mathrm{sys}}$) as well as microwave ($\zeta_\mathrm{m}$) and optical ($\zeta_\mathrm{o}$) extraction efficiencies. Accordingly, the transduction process is characterized by the pair of parameters [$\eta_\mathrm{t}$, $\bar{n}$], as noted in Fig.~\ref{fig:networksetup}. In evaluating the thermal noise contribution of the transducer, we assume a representative microwave frequency of $8~\mathrm{GHz}$; the corresponding thermal occupation at optical frequencies is negligible. Realistic values of $\bar{n}$ used in the subsequent performance evaluation are discussed in Appendix~\ref{app:nbar}. 


Using the displacement-operator approach, the state after the transduction channel is given by~\cite{GaussianChannel}:
\begin{equation}
\label{eq:transduction}
\rho_R^{\mathrm{tr}}(\xi) = \rho_R^{\mathrm{ch}}\Bigl(\sqrt{\eta_\mathrm{t}}\,\xi\Bigr)e^{-N_\mathrm{t}\,|\xi|^2}\,.
\end{equation}

\subsection{Coupling Mechanism}
\label{sec:coupling}
The next step involves evaluating the coupling mechanism between the microwave photon output from the transducer and a two-level superconducting matter qubit of subsystem $M$ initialized in the ground state ($M_A$ for Alice's segment and $M_B$ for Bob's segment). The eigenstates in the bare basis are $\{\ket{g}, \ket{e}\}$. This process ideally performs a swap of the flying qubit state to the matter qubit, according to the transformations: $\ket{g\,0} \rightarrow \ket{g\,0}$, $\ket{g\,1} \rightarrow \ket{e\,0}$. This dynamics is described by circuit quantum electrodynamics (CQED) notation, in which a resonator is capacitively coupled to a superconducting qubit (e.g., a transmon qubit) \cite{QED}. On resonance, i.e., when the qubit frequency matches the resonator frequency (zero detuning)~\cite{resonance}, the energy exchange between the two systems gives rise to Rabi oscillations whereby a single photon is transferred between the matter qubit and resonator with unit probability. This process has been experimentally realized in several works using an external tunable coupler based on a superconducting quantum interference device (SQUID)~\cite{coupling1,coupling2}. The dynamics of this interaction can be approximated within the rotating-wave approximation (RWA) by the Jaynes–Cummings interaction Hamiltonian~\cite{couplinghamiltonian}:
\begin{equation}
H_{\mathrm{int}}
= \hbar\,g_r\bigl(a^\dagger \,\sigma^-\;+\;a\,\sigma^+\bigr),
\end{equation}
where $a^\dagger$ and $a$ denote the microwave photon creation and annihilation operators, respectively, $\sigma^+ = \ket{e}\bra{g}$ and $\sigma^- = \ket{g}\bra{e}$ are the qubit raising and lowering operators, respectively, $g_r$ is the vacuum Rabi coupling strength (rad/s), and $\hbar$ is the reduced Planck's Constant.  
After a coupling time $t=\tfrac{\pi}{2g_r}$, the input state $\ket{g\,1}$ evolves to $-i\ket{e\,0}$, while $\ket{g\,0}$ remains unchanged~\cite{resonance}.

We propose the following joint Kraus operators to analytically model the imperfect coupling process:
\begin{equation}
\begin{split}  
    &K_0 = |g \, 0\rangle_{MT}\langle g \, 0|_{MT} - i \sqrt{\eta_{\rm qb} \eta_{\rm ph}} |e\,0\rangle_{MT}\langle g\,1|_{MT}, \\  
    &K_1 = \sqrt{1 - \eta_{\rm qb} \eta_{\rm ph}} |g\,0\rangle_{MT}\langle g \, 1|_{MT},  
\end{split}
\end{equation}
where the efficiency parameters are defined as follows:
   \begin{itemize}
       \item $\eta_{\rm ph}$: probability that the microwave photon is not lost during the coupling interaction with the qubit;
       \item $\eta_{\rm qb}$: probability that the qubit makes the correct transition to the excited state upon absorbing the photon, conditional on the photon being available.
   \end{itemize}
Starting from the reduced state in Eq.~(\ref{eq:transduction}), the full density matrix is obtained using the Glauber formula~\cite{olivares_introduction_2021}  
\(\rho_{TR}^{\text{tr}} = \frac{1}{\pi} \int \rho_R^{\text{tr}}(\xi) D_T^\dagger(\xi) \, d^2\xi \).  
Finally, including the superconducting subsystem $M$ into the post-transduction global state \(\rho_{MTR}^{\mathrm{tr}} = |g\rangle_M\langle g|_M\otimes\rho_{TR}^{\mathrm{tr}}\), we obtain the full state description: 
\begin{align}
    \rho_{MTR}^{\mathrm{coupl}} 
    = \sum_{i=0,1} \bigl(K_i \otimes \mathbb{I}_R \bigr) \,\rho_{MTR}^{\mathrm{tr}}\, \bigl(K_i^\dagger \otimes \mathbb{I}_R\bigr),
\end{align}
where $\mathbb{I}_R$ is the identity operator that acts on subsystem $R$.
As shown in Appendix~\ref{app:coupling-formula}, the photonic subsystem $T$ in $\rho_{MTR}^{\mathrm{coupl}}$ is disentangled from subsystems $M$ and $R$, and can therefore be traced out and neglected in the subsequent analysis.

\subsection{Swapping Circuit}
\label{sec:swapping}
Recalling the symmetry of the two links Alice–TESQR and Bob–TESQR, we can write $\rho_{MR}^{\mathrm{coupl}}=\rho_{M_AR_A}^{\mathrm{coupl}}=\rho_{M_BR_B}^{\mathrm{coupl}}$ and obtain the full system density matrix as:
\begin{equation}
\label{eq:rho0text}
\rho_0=\rho_{M_AR_AM_BR_B}^{\mathrm{coupl}}=\rho_{M_AR_A}^{\mathrm{coupl}}\otimes\rho_{M_BR_B}^{\mathrm{coupl}}. 
\end{equation}

The TESQR's QPU applies the gate-based version of the BSM \cite{entanglswapping} on its two superconducting qubits, \(M_A\) and \(M_B\).
We model noisy gate operations through depolarizing noise~\cite{QRenconding,PRXQuantumQR}. Consequently, the transformations (for application of the CNOT unitary \(U_{\rm C}\) on qubits $i$ and $j$, and of the Hadamard unitary \(U_\mathrm{H}\) on qubit $i$) are expressed in our model as:
\begin{equation}
\label{eq:Ucnot}
\begin{split}
U_{\mathrm{C}}^{(ij)}\,\rho_0\,U_{\mathrm{C}}^{(ij)\dagger}
&\longrightarrow
\rho_1=(1-p_\mathrm{CNOT})\,U_{\mathrm{C}}^{(ij)}\,\rho_0\,U_{\mathrm{C}}^{(ij)\dagger} \\
&\quad
+ \frac{p_\mathrm{CNOT}}{4}\,\mathrm{Tr}_{ij}[\rho_0]\otimes \mathbb{I}_{ij},
\end{split}
\end{equation}
\begin{equation}
\begin{split}
U_\mathrm{H}^{(i)}\,\rho_1\,U_\mathrm{H}^{(i)\dagger}
&\longrightarrow
\rho_2=(1-p_\mathrm{H})\,U_\mathrm{H}^{(i)}\,\rho_1\,U_\mathrm{H}^{(i)\dagger} \\
&\quad
+ \frac{p_\mathrm{H}}{2}\,\mathrm{Tr}_{i}[\rho_1]\otimes \mathbb{I}_{i},
\end{split}
\end{equation}
where \( \mathrm{Tr}_{ij}\) and \(\mathrm{Tr}_{i}\) denote the partial trace over, respectively, subsystems $i,j$ and subsystem $i$; while $ \mathbb{I}_{ij}$ and $\mathbb{I}_{i}$ are the identity operators on the corresponding subsystems. $p_\mathrm{CNOT}$ and $p_\mathrm{H}$ represent the gate error probabilities for the CNOT and Hadamard gates, respectively, as indicated in Fig. \ref{fig:networksetup}. In reference to Eq.~(\ref{eq:rho0text}), $i=M_A$ and $j=M_B$.

Finally, we describe the imperfect single-qubit measurements on both $M_A$ and $M_B$ in 
the computational basis by a Positive Operator-Valued Measure (POVM)~\cite{QRpurification} with the corresponding imperfect measurement operators for outcomes $\ket{g}$ and $\ket{e}$ defined, respectively, as:
\begin{equation}
\label{eq:proj}
\begin{split}
     &P_g = (1 - \varepsilon) |g\rangle_M\langle g|_M + \varepsilon |e\rangle_M\langle e|_M,\\
     &P_e = \varepsilon |g\rangle_M\langle g|_M + (1 - \varepsilon) |e\rangle_M\langle e|_M,
     \end{split}
\end{equation}
where, $\varepsilon$ quantifies the quality of the measurement on the computational basis.

\section{Performance Evaluation}
\label{sec:performanceval}

To evaluate the performance of the proposed scheme, it is necessary to first define the ideal entangled state of the remote matter qubits $R_A$ and $R_B$. Assuming the absence of noise and losses during channel transmission, transduction, and coupling processes (i.e., $\eta_\mathrm{c}=1$, $\eta_\mathrm{t}=1$, $\bar{n}=0$, $\eta_\mathrm{ph}=1$, $\eta_\mathrm{qb}=1$), the ideal state after coupling is given by:
\begin{equation}
\begin{split}
\ket{\psi}_{R_AM_AR_BM_B}&=\left[ \frac{1}{\sqrt{2}} \left( |gg\rangle_{R_AM_A} - i |ee\rangle_{R_AM_A} \right) \right] \\ &\otimes \left[ \frac{1}{\sqrt{2}} \left( |gg\rangle_{R_BM_B} - i |ee\rangle_{R_BM_B} \right) \right].
\end{split}
\end{equation}

During the subsequent swapping phase, the ideal CNOT ($p_{\mathrm{CNOT}}=0$) and Hadamard ($p_\mathrm{H}=0$) gates, together with the ideal POVM measurements ($\varepsilon=0$) on qubits $M_A$ and $M_B$, project the remote qubits $R_A$ and $R_B$ into an entangled state $\ket{\psi^{\mathrm{ideal}}_{R_AR_B}}$. Its explicit form follows from Eqs.~(\ref{eq:Ucnot})--(\ref{eq:proj}) and is summarized in Table~\ref{tab:entangled_states}.

\begin{table}[htb]
\centering
\caption{Ideal entangled states of remote qubits $R_A$ and $R_B$, 
conditioned on the measurement outcome of $M_A M_B$.}
\renewcommand{\arraystretch}{1.7} 
\begin{ruledtabular}
\begin{tabular}{c  c}
Outcome $(M_AM_B)$ & Ideal post-measurement state $|\psi_{R_AR_B}^{\mathrm{ideal}}\rangle$ \\ 
\hline
$|g g\rangle_{M_AM_B}$ & $\tfrac{1}{\sqrt{2}}(|g g\rangle_{R_AR_B} - |e e\rangle_{R_AR_B})$ \\
$|e g\rangle_{M_AM_B}$ & $\tfrac{1}{\sqrt{2}}(|g g\rangle_{R_AR_B} + |e e\rangle_{R_AR_B})$ \\
$|g e\rangle_{M_AM_B}$ & $-\,i\,\tfrac{1}{\sqrt{2}}(|g e\rangle_{R_AR_B} + |e g\rangle_{R_AR_B})$ \\
$|e e\rangle_{M_AM_B}$ & $\,i\,\tfrac{1}{\sqrt{2}}(-|g e\rangle_{R_AR_B} + |e g\rangle_{R_AR_B})$ \\
\end{tabular}
\label{tab:entangled_states}
\end{ruledtabular}
\end{table}

\subsection{Output Fidelity}
\label{sec:outputfidelity}
For a given measurement result of $M_AM_B$, let $\rho_{R_AR_B}^{\mathrm{final}}$ denote the $R_AR_B$ density matrix and let $|\psi_{R_AR_B}^{\mathrm{ideal}}\rangle$ be the associated ideal target state. Accordingly, $\rho_{R_AR_B}^{\mathrm{final}}$ denotes the general scenario with nonzero network noise parameters and is analytically obtained by applying the noise operation model introduced in Section~\ref{sec:model}. The quality of the entangled state generated between qubits $R_A$ and $R_B$ is quantified by computing its fidelity with respect to the corresponding ideal state reported in Table~\ref{tab:entangled_states}. The final 
fidelity is defined as:
\begin{equation}
\label{eq:Ffinal}
F_{\mathrm{final}} = \langle \psi_{R_AR_B}^{\mathrm{ideal}} \,|\, 
      \rho_{R_AR_B}^{\mathrm{final}} \,|\,
      \psi_{R_AR_B}^{\mathrm{ideal}} \rangle,
\end{equation}
which measures the overlap between the experimentally obtained state and the ideal entangled state corresponding to the specific measurement outcome. Indeed, fidelity is a straightforward yet crucial performance metric for assessing quantum-state quality, as maintaining high entanglement fidelity directly impacts the performance and security of quantum applications such as key distribution and teleportation~\cite{tcom}.

\begin{figure*}[htb]
    \centering
    \includegraphics[width=\linewidth]{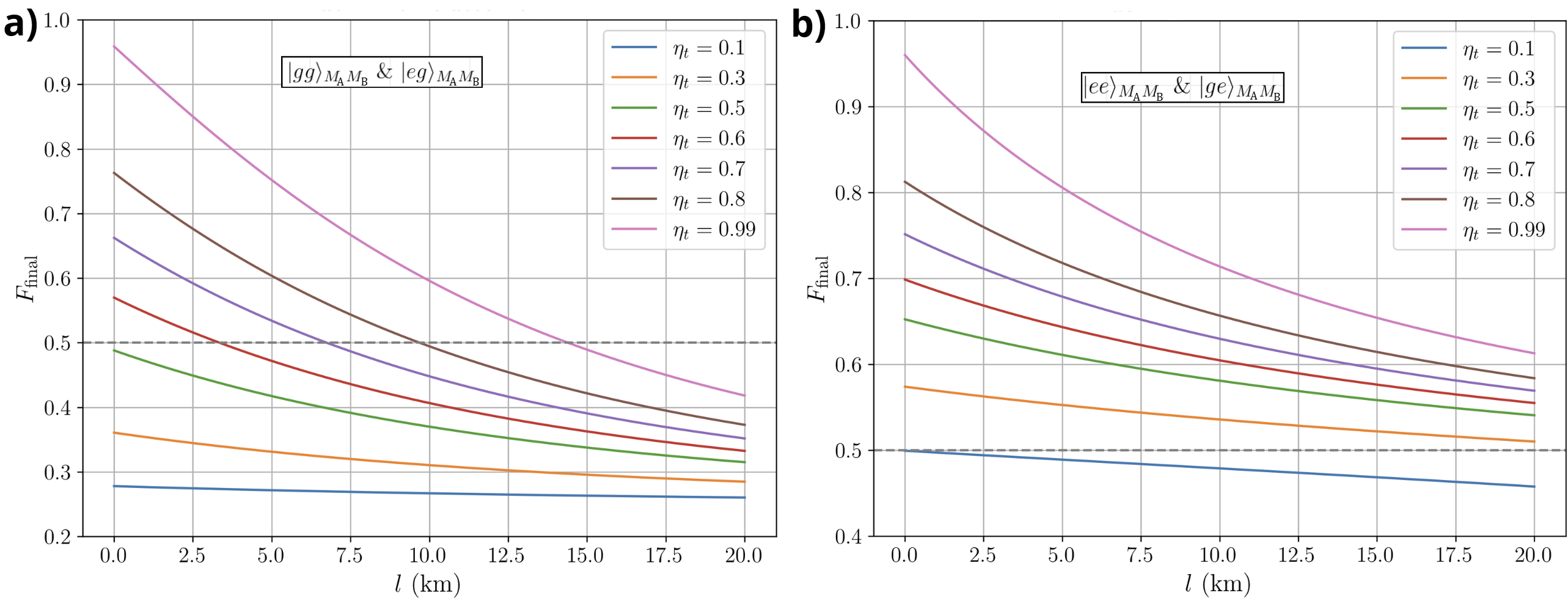}
    \caption{\mhl{Final output fidelity $F_{\mathrm{final}}$ as a function of fiber length $l$ (km), for various transduction efficiency values and for $M_AM_B$ measurement outcomes: a) $\ket{gg}_{M_AM_B}$ and $\ket{eg}_{M_AM_B}$, b) $\ket{ee}_{M_AM_B}$ and $\ket{ge}_{M_AM_B}$. The other network parameters are set according to Table~\ref{tab:simulation_parameters}.}}
    \label{fig:fidelityvsl}
\end{figure*}

\begin{figure*}[htb]
    \centering
    \includegraphics[width=\linewidth]{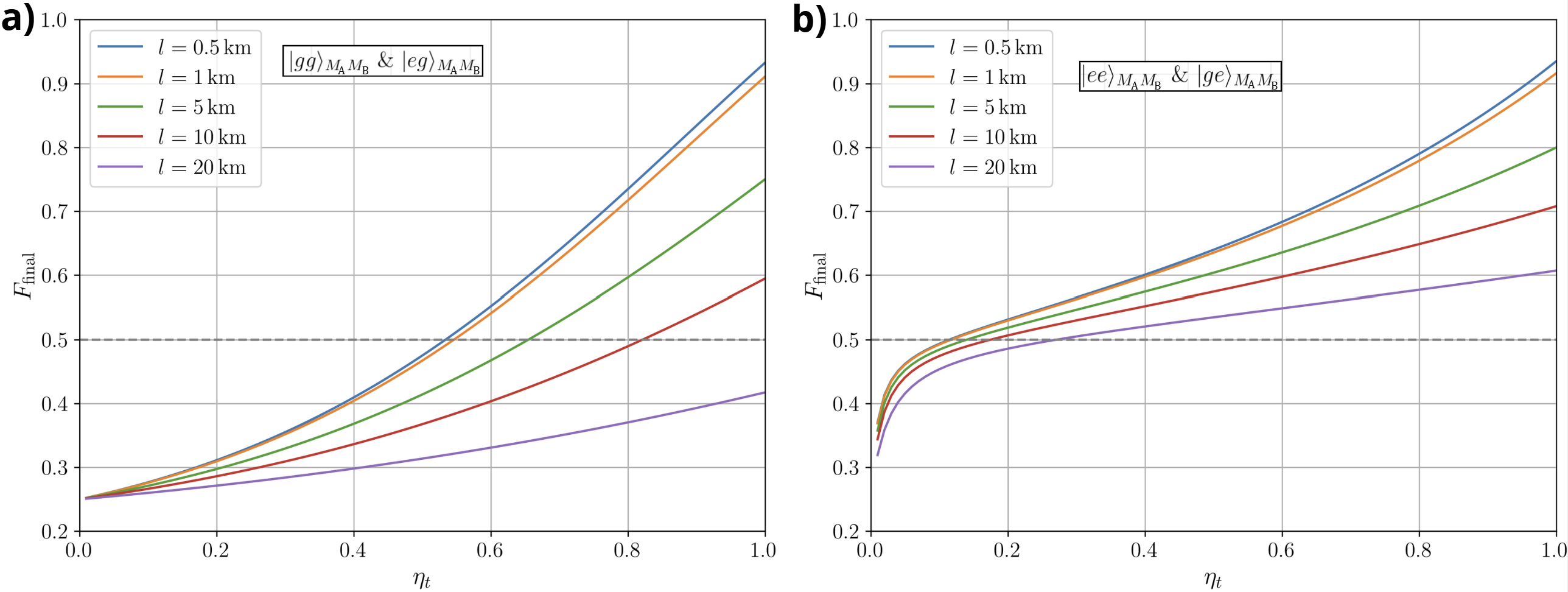}
    \caption{\mhl{Final output fidelity $F_{\mathrm{final}}$ as a function of transduction efficiency $\eta_\mathrm{t}$, for various fiber length values and for $M_AM_B$ measurement outcomes: a) $\ket{gg}_{M_AM_B}$ and $\ket{eg}_{M_AM_B}$, b) $\ket{ee}_{M_AM_B}$ and $\ket{ge}_{M_AM_B}$. The other network parameters are set according to Table \ref{tab:simulation_parameters}.}}
    \label{fig:fidelityvsetat}
\end{figure*}

The typical values of the relevant network architecture parameters are reported in Table ~\ref{tab:simulation_parameters} \mhl{\cite{Rueda2019,li_error_2023,IBM,engineeringguide,kubo_fast_2023}}. In Fig.~\ref{fig:fidelityvsl}, we show the behavior of $F_{\mathrm{final}}$ as a function of the length of a single source--TESQR segment, hence, the total distance between Alice and Bob corresponds to twice the value reported.
Concerning the values of $\bar{n}$, we consider the setup analyzed in Appendix~\ref{app:nbar}, with a temperature $T_\mathrm{sys}= 0.1$~K. The curves are computed for different values of the transduction efficiency, and they are computed analytically according to the model developed in Sec.~\ref{sec:model}. 
Similarly, Fig.~\ref{fig:fidelityvsetat} illustrates the corresponding fidelity trends as a function of the transduction efficiency for several fiber length values.
\begin{table}[htb]
\centering
\caption{Numerical values of the network model parameters used for performance evaluation.}
\renewcommand{\arraystretch}{1.1}
\begin{ruledtabular}
\begin{tabular}{c c}
Parameter & Value \\ 
\hline
$l_{\mathrm{att}}$ & 22 km \\
$T_{\mathrm{sys}}$ & 0.1 K \\
$\zeta_{\mathrm{m}}$ & 0.9 \\
$\zeta_{\mathrm{o}}$ & 0.9 \\
$\eta_{\mathrm{ph}}$ & 0.99 \\
$\eta_{\mathrm{qb}}$ & 0.99 \\
$p_{\mathrm{H}}$ & $2\times10^{-4}$ \\
$p_{\mathrm{CNOT}}$ & $2\times10^{-3}$ \\
$\varepsilon$ & $5\times10^{-3}$ \\
\end{tabular}
\end{ruledtabular}
\label{tab:simulation_parameters}
\end{table}

Several features can be observed from Figs.~\ref{fig:fidelityvsl} and \ref{fig:fidelityvsetat}. 
The fidelity shows a symmetric trend between $\ket{gg}_{M_AM_B}$ and $\ket{eg}_{M_AM_B}$, and $\ket{ge}_{M_AM_B}$ and $\ket{ee}_{M_AM_B}$. 
This is consistent with the corresponding symmetry of the ideal post-measurement states, reported in Table~\ref{tab:entangled_states}. As a result of fiber photon loss, transduction thermal noise, and imperfect coupling, qubits $M_A$ and $M_B$ 
enter the swapping procedure in the ground state. This, when followed by the CNOT and the Hadamard gates, effectively leads to the Bell-measurement output combinations $\ket{gg}_{M_AM_B}$ and $\ket{eg}_{M_AM_B}$.

In addition, all cases exhibit a common exponential decay of the fidelity with increasing source--TESQR segment length, due to the exponential attenuation of the optical fiber. 
However, the outcomes $\ket{ge}_{M_AM_B}$ and $\ket{ee}_{M_AM_B}$ 
display a more favorable behavior compared to the other two cases ($\ket{gg}_{M_AM_B}$ and $\ket{eg}_{M_AM_B}$). 
In particular, Fig.~\ref{fig:fidelityvsetat}(b) shows that a transduction efficiency above $0.4$ is sufficient to reach a fidelity greater than $0.5$ even for $l = 20$~km (corresponding to a total Alice--Bob separation of $40$~km).
Indeed, the value $F_{\mathrm{final}}=0.5$ represents the minimum threshold required to certify the presence of entanglement \cite{minimumfidelity}, which is indicated in Figs.~\ref{fig:fidelityvsl} and \ref{fig:fidelityvsetat} by a horizontal gray dashed line.

The other two outcomes ($\ket{gg}_{M_AM_B}$ and $\ket{eg}_{M_AM_B}$) correspond to cases in which noise contributions—primarily due to photon loss and transduction noise—are more pronounced. 
This is confirmed by the fact that, with transduction efficiency approaching unity and short distances, i.e., negligible photon losses from the communication link, the fidelities associated with the four measurement outcomes approach the same values, as shown in Fig.~\ref{fig:fidelityvsetat}, in the upper-right region. 
Conversely, in the low-efficiency and long-distance regime, the fidelity trend corresponding to $\ket{gg}_{M_A M_B}$ and $\ket{eg}_{M_A M_B}$ outcomes diverge maximally from those associated with $\ket{ge}_{M_A M_B}$ and $\ket{ee}_{M_A M_B}$.
For a link distance of $10$~km, the fidelity trend shown in Fig. \ref{fig:fidelityvsetat}(a) surpasses the $0.5$ threshold only for $\eta_t \geq 0.83$, reaching a maximum of $F_{\mathrm{final}}=0.6$ at ideal unit transduction efficiency. 
By contrast, in Fig. \ref{fig:fidelityvsetat}(b), the fidelity already exceeds the $0.5$ threshold for $\eta_t \geq 0.25$, eventually attaining a maximum value of $0.71$.

\subsection{Probability of Success}
\label{sec:probofsuccess}
\begin{figure}[htb]
    \centering
    \includegraphics[width=\linewidth]{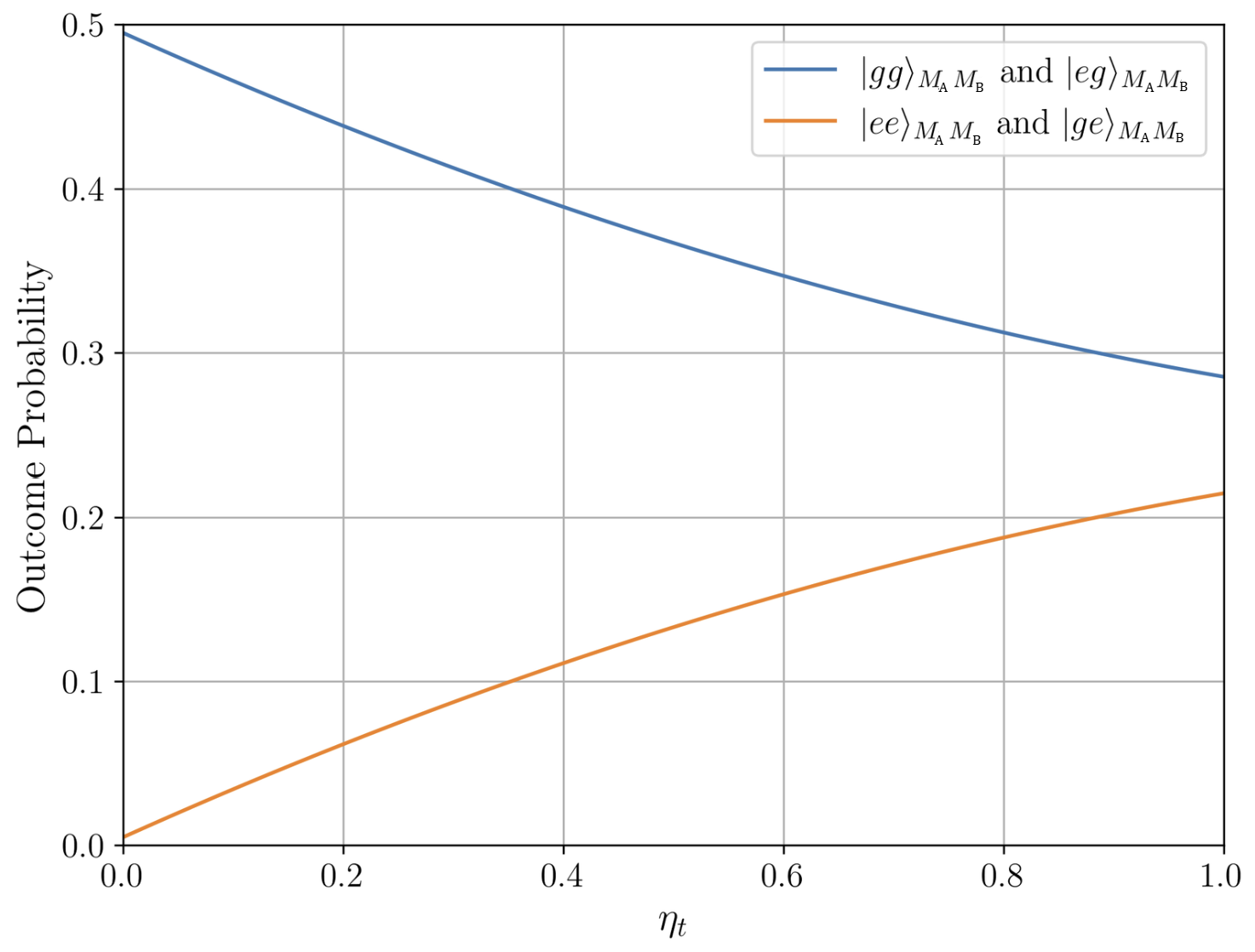}
    \caption{Measurement outcome probabilities for $M_A$ and $M_B$ superconducting qubits as a function of transduction efficiency $\eta_\mathrm{t}$, with $l=10$ km and other network parameters set according to Table \ref{tab:simulation_parameters}. }
    \label{fig:probability}
\end{figure}

Figure~\ref{fig:probability} shows the outcome measurement probabilities of $M_AM_B$ qubits as a function of the transduction efficiency, for a fiber length $l=10$ km and the parameters listed in Table~\ref{tab:simulation_parameters}.

Figure~\ref{fig:probability} confirms the symmetry of the possible output states and, in the considered scenario, the higher occurrence probability of the states $\ket{gg}_{M_AM_B}$ and $\ket{eg}_{M_AM_B}$. 
Moreover, as $\eta_t \rightarrow 1$, the probabilities redistribute and approach the equiprobable value of $0.25$, which would ideally be reached in a noiseless environment. 

To assess the protocol's probability of success, we note that the hybrid TESQR scheme \mhl{always} produces a final state shared by qubits $R_A$ and $R_B$. As shown in Fig.~\ref{fig:probability}, noise redistributes the outcome probabilities while preserving their total sum. Hence, a measurement result is always obtained, although the fidelity of the corresponding state may be reduced by losses and imperfections. In contrast, optical BSM schemes~\cite{entangling,entanlgementtrapped} are intrinsically probabilistic and therefore do not always yield an outcome—for instance, this happens 
due to photon loss, imperfect interference, or indistinguishability~\cite{entanlgementtrapped,saha_high-fidelity_2025,opticalQRchallenge}. Such events must therefore be discarded. The TESQR approach, instead, always provides a measurement outcome. As a result, even noisy states remain available for further processing, including entanglement purification as discussed below. In this sense, the protocol is \mhl{operationally} deterministic, as it eliminates the probabilistic nature of optical BSMs.

Building on this distinction, it is convenient to introduce a quantitative measure of success that captures the usefulness of the outcomes produced by the TESQR scheme, hence allowing for a comparison with the fully photonic swapping protocol. In particular, 
we consider an output state $R_AR_B$ to be entangled and thus useful for further processing only if its fidelity exceeds the threshold of $0.5$~\cite{minimumfidelity,entangling}. Accordingly, in our network scenario, where we deal with four possible measurement outcomes on $M_A$ and $M_B$, we define the probability of success as:
\begin{equation}
\label{eq:Probsuccess}
    P_{\mathrm{success}} = \sum_{i,j \in \{\ket{g},\ket{e}\}} p_{i,j} \, \Theta\!\left(F_{\mathrm{final}}^{i,j} - 0.5\right),
\end{equation}
where $p_{i,j}$ denotes the probability of obtaining the $M_AM_B$ measurement outcome corresponding to $\ket{ij}_{M_AM_B}$, $F_{\mathrm{final}}^{i,j}$ represents the corresponding output fidelity with respect to the ideal Bell state $\ket{\psi_{R_AR_B}^{\mathrm{ideal}}}$ (as defined in Eq.~(\ref{eq:Ffinal})), and $\Theta(\cdot)$ is the Heaviside step function ($\Theta(x) = 0 \text{ if } x < 0, \, \Theta(x) = 1 \text{ if } x \ge 0$). \mhl{Under this definition, an outcome counts as successful only if its individual fidelity exceeds the 0.5 threshold.} 

\begin{figure}[htb]
    \centering
    \includegraphics[width=\linewidth]{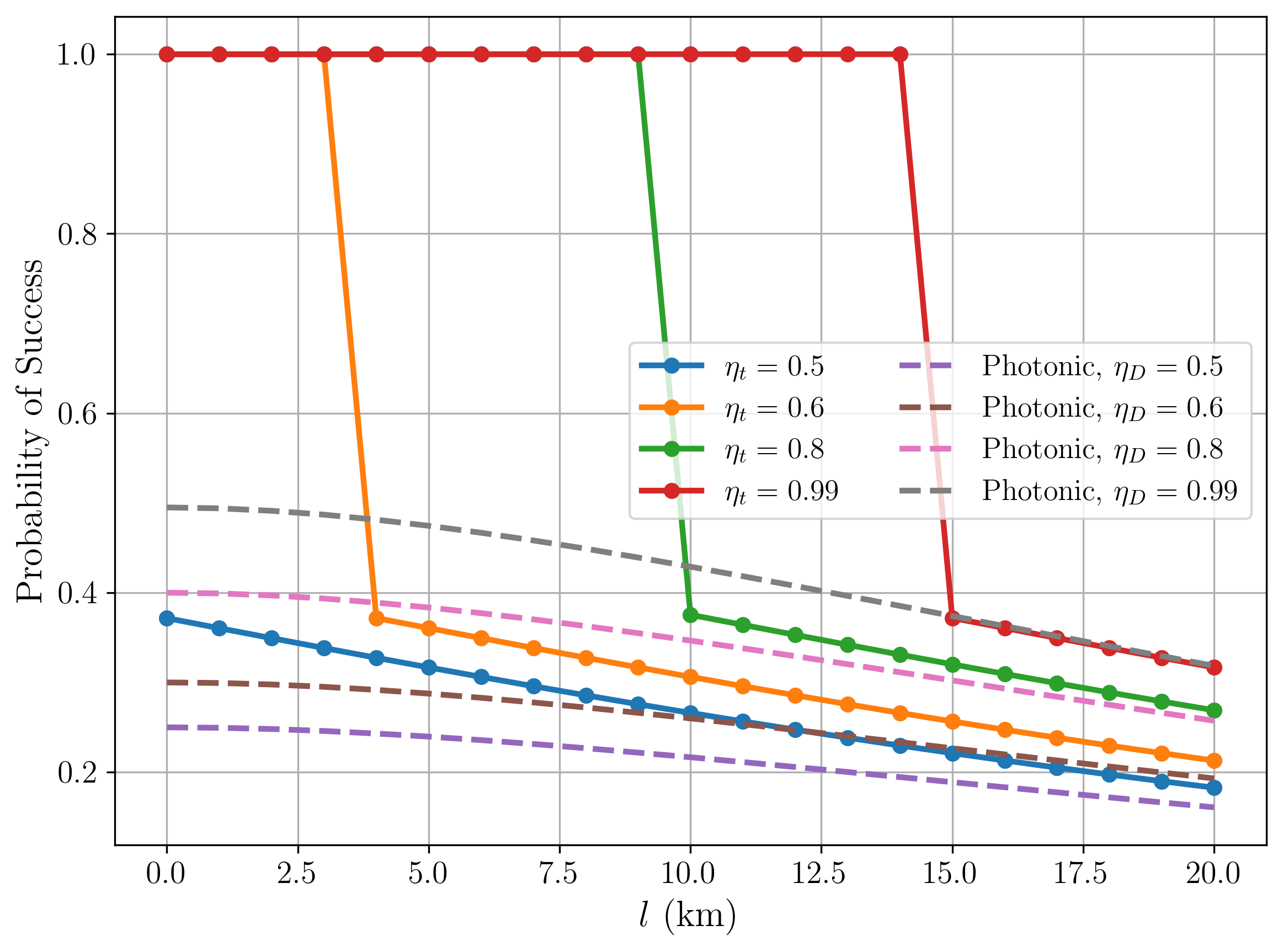}
    \caption{Swapping success probability versus fiber length $l$ (km) for the hybrid TESQR scheme (solid curves) and the photonic counterpart (dashed curves), plotted for different transduction efficiencies $\eta_\mathrm{t}$ and photodetection efficiencies $\eta_\mathrm{D}$.}
    \label{fig:probabilityofsuccess}
\end{figure}

As a benchmark for our approach, we consider the equivalent optical BSM in a single-rail encoding scheme, consisting of a balanced beamsplitter and two number-resolving photodetectors~\cite{entangling}. In this setup, a successful heralded swapping event corresponds to the detection patterns $[0,1]$ and $[1,0]$, where exactly one of the photodetectors is triggered by a single photon. As the only source of noise in this scheme, we account for the detection efficiency $\eta_\mathrm{D}$ of the photodetectors. Accordingly, in Fig.~\ref{fig:probabilityofsuccess} we compare the swapping success probability of the proposed hybrid TESQR scheme, $P_\mathrm{success}$ [Eq.~(\ref{eq:Probsuccess})] (solid curves), with that of the photonic counterpart (dashed curves)~\cite{entangling}, as a function of the fiber length $l$, and for different values of $\eta_\mathrm{t}$ and $\eta_\mathrm{D}$, respectively. \mhl{Unlike the purely photonic case, and following a worst-case analysis approach, all other noise parameters listed in Table~\ref{tab:simulation_parameters} are taken into account in the TESQR scheme.}

Figure~\ref{fig:probabilityofsuccess} shows that, for equal values of transduction efficiency and detection efficiency, the TESQR scheme consistently achieves a higher probability of success, and thus a larger expected entangling rate, compared to the purely photonic counterpart. In particular, for $\eta_\mathrm{t} \gtrsim 0.6$, there exists an initial fiber range where $P_{\mathrm{success}}$ is equal to unity for the proposed scheme. In this regime, all possible measurement outcomes of $M_A$ and $M_B$ yield to entangled states of $R_A$ and $R_B$ with fidelity above $0.5$. \mhl{Within this parameter range, entanglement distribution can therefore be regarded as deterministic.} The extent of this deterministic region increases with $\eta_\mathrm{t}$, reaching a limit distance of approximately $14\,\mathrm{km}$ for $\eta_\mathrm{t}=0.99$. Beyond this flat region, $P_{\mathrm{success}}$ exhibits a drop, which arises because, above this distance threshold, only two of the four possible measurement outcomes yield fidelities above $0.5$. Nevertheless, even after this drop, the TESQR curves remain consistently above those of the photonic scheme. The performance gap further widens as the corresponding transduction or detection efficiency decreases. \mhl{We highlight that the success probability of the photonic scheme is fundamentally bounded above by $0.5$, in agreement with the ideal theoretical limit~\cite{calsamiglia_maximum_2001}.} Finally, the TESQR approach exhibits an average and maximum improvement of 63\% and 159\%, respectively, in the entanglement distribution success probability compared to the photonic-only counterpart across the tested scenarios.
\rhl{We emphasize that the reported percentage gains quantify the rate of generating an entangled resource suitable for further processing, rather than the quality of the resulting entangled states; therefore, the comparison with the photonic scheme should be interpreted in terms of the rate of usable entanglement generation, not output-state fidelity.}


\section{Purification Enhancement}
\label{sec:simresults}
Entanglement purification is a well-established method to enhance the fidelity of output states, at the expense of additional quantum state processing at the network nodes. In this regard, our hybrid solution with TESQR mitigates this issue by operating with superconducting qubits, thereby enabling a direct circuit-based implementation. By contrast, in photonic schemes entanglement purification can be realized, for instance, through recurrence protocols \cite{pan_entanglement_2001}, cross-Kerr nonlinearities, hyperentanglement \cite{huang_experimental_2022}, or multipartite entangled states \cite{Yan:21}. However, such approaches are generally highly resource-demanding 
\cite{yan_advances_2023}.

\begin{figure*}[htb]
    \centering
    \includegraphics[width=0.75\linewidth]{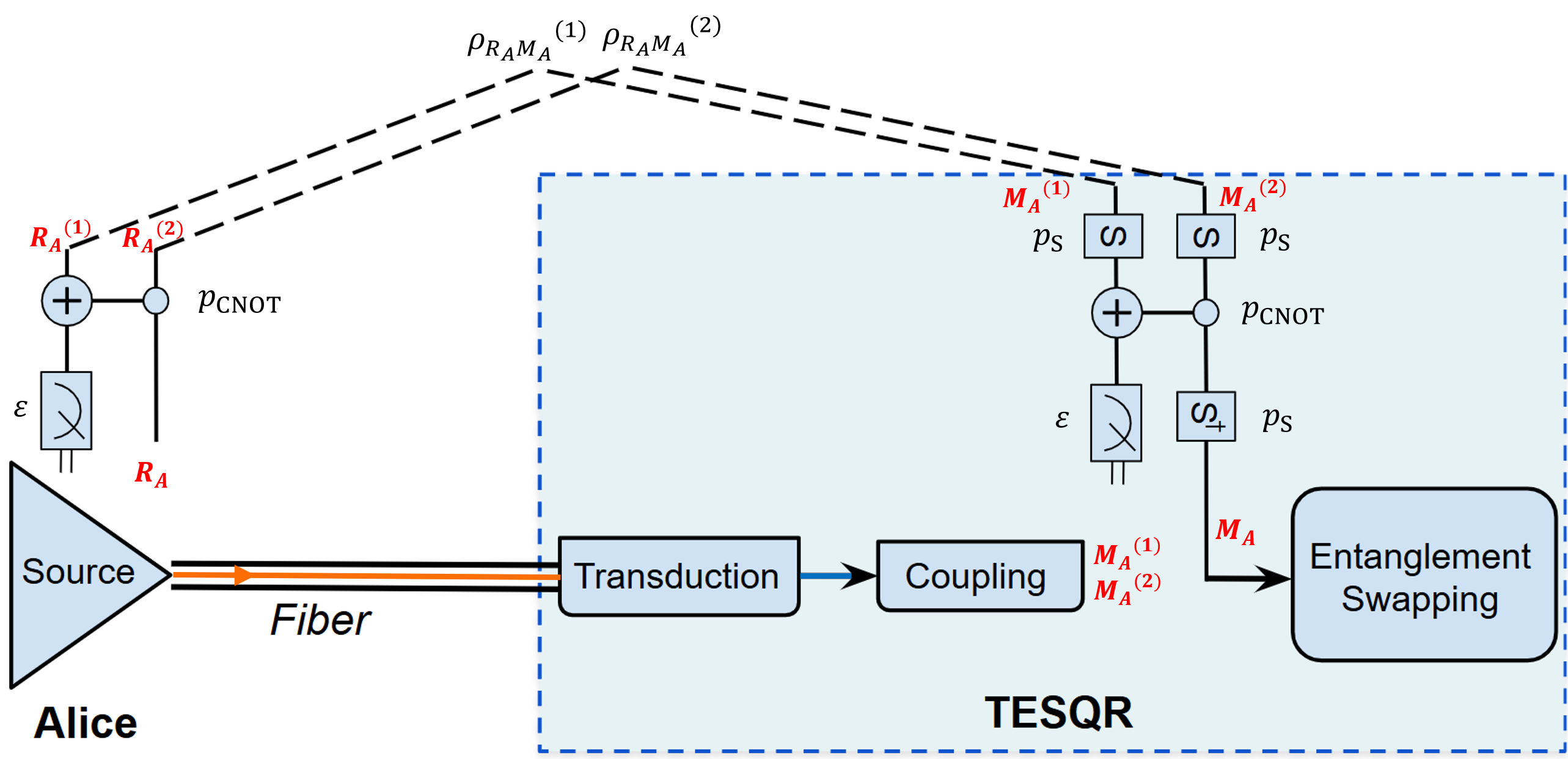}
    \caption{Alice–TESQR segment incorporating single-round BBPSSW entanglement purification prior to swapping. Two 
    instances of the source state undergo noisy fiber transmission, transduction, and coupling, producing the noisy states ${\rho_{R_AM_A}}^{(1)}$ and ${\rho_{R_AM_A}}^{(2)}$. TESQR applies local $S$ gates to $M_A^{(1,2)}$ to obtain Werner form. Bilateral CNOTs (at both Alice and TESQR) and measurements on the target qubits follow. Success is heralded if the outcomes match. In that case, an $S^\dagger$ gate is applied to the surviving TESQR qubit before the swapping procedure (Fig.~2). The scheme is symmetric for the Bob–TESQR segment. \mhl{Optical and microwave photon paths are highlighted in orange and blue, respectively.}}
    \label{fig:setuppurifcation}
\end{figure*}

In this scenario, we are primarily interested in assessing the performance contribution of the TESQR node, and given that the swapping scheme introduces additional noise, it is advantageous to apply entanglement purification immediately after the coupling process. This is performed independently on the two segments Alice–TESQR and Bob–TESQR, acting on the pairs $\rho_{M_AR_A}$ and $\rho_{M_BR_B}$. In each segment, the sequence of operations is: TESQR $\rightarrow$ entanglement purification $\rightarrow$ entanglement swapping. For the purification scheme, we adopt the classical BBPSSW (Bennett–Brassard–Popescu–Schumacher–Smolin–Wootters) protocol \cite{Bennet}. Figure~\ref{fig:setuppurifcation} illustrates the single-round protocol applied to our network model with reference to the Alice–TESQR segment. Alice (Bob) generates two instances of the source state $\rho_{TR}^{\mathrm{source}}$, which are subsequently processed in the photonic domain as described earlier, eventually yielding, after coupling, two instances ${\rho_{R_AM_A}}^{(1)}$ and ${\rho_{R_AM_A}}^{(2)}$. 

We first transform the states into a Werner-like form, as required by the original protocol \cite{Bennet}. In our scenario, the TESQR node applies a local $S$ gate to both superconducting matter qubits $M_A^{(1)}$ and $M_A^{(2)}$ (see Appendix~\ref{app:Spurifica}). Subsequently, both parties perform a bilateral CNOT gate followed by computational-basis POVM measurements on the target qubits. The purification step is deemed successful when the measurement outcomes coincide. In that case, the TESQR node applies an adjoint $S$ gate ($S^\dagger$) to the remaining superconducting qubit, thereby restoring it to its original form. This purified qubit is then used in the entanglement swapping procedure (Fig.~\ref{fig:networksetup}). 

Consistent with our previous noise modeling, the $S$ gates are characterized by a depolarizing channel noise acting with probability $p_\mathrm{S}$. To ensure a realistic performance evaluation, the other purification operations (CNOT gates and measurements) are also modeled with the noise parameters described in Section~\ref{sec:swapping}.

\subsection{Simulation Framework}
\label{sec:simulationframework}

To numerically assess the fidelity after  purification, we implement a Monte Carlo simulator based on QuTiP library~\cite{Qutip}. In this framework, we implement the building blocks of the proposed hybrid network architecture. To ensure reproducibility, the software code is available at~\cite{simulatorTESQR}. Fiber propagation is described as an amplitude-damping (pure-loss) channel acting on the photonic Fock space. Kraus operators for the lossy photonic mode are applied probabilistically to state vectors in order to generate stochastic trajectories (implemented via sampling over Kraus outcomes). 
\mhl{In the simulation environment, the photonic Hilbert space is truncated to a finite Fock basis in order to balance modeling accuracy and computational complexity. This approximation does not significantly affect the subsequent modeling of the transduction process, as further discussed in Appendix~\ref{app:dim_photonic}.}

Electro-optic quantum transduction is modeled as a beamsplitter-type unitary interaction between the system mode and a thermal environment, 
\(
U(\theta)=\exp\{\theta(a^\dagger b - a b^\dagger)\},
\) where $a$ and $a^\dagger$ are, respectively, the lowering and raising operator for the input signal state (and $b$, $b^\dagger$ are the equivalent version for the environment state), and $\theta=arccos(\sqrt{\eta_t})$. The environment is initialized in a thermal bath state \(\rho_{\rm th}(\bar{n})\) \cite{Nielsen_Chuang_2010}. Importantly, the photon conversion 
is implemented at the density-matrix level: the joint system–environment state evolves under the unitary, after which a partial trace over the environment is performed to obtain the reduced system state. This realizes the Gaussian-loss model, i.e., the beamsplitter with added noise~\cite{Jing2}. From the resulting reduced density matrix, we then sample a pure state via eigen-decomposition, that enables straightforward state-vector propagation in the subsequent simulation steps. 

Photonic–matter qubit coupling is modeled with the appropriate set of previously introduced Kraus elements, which are applied probabilistically to the input state vector. Gate imperfections (in both purification and swapping) are described as local depolarizing channels, with Kraus set 
\(\{\sqrt{1-p}\,I,\sqrt{p/3}\,X,\sqrt{p/3}\,Y,\sqrt{p/3}\,Z\}\) applied to the affected qubit(s) \cite{Nielsen_Chuang_2010}. Imperfect POVM measurements are incorporated by probabilistic Kraus sampling, parametrized by the measurement error rate $\varepsilon$. Finally, the post-swap reduced state of the remote qubits is retained in density-matrix form and used for fidelity evaluation. 

\subsection{Results and Discussion}
\label{sec:resultsanddiscussion}
\begin{figure*}[htb]
    \centering
    \includegraphics[width=\linewidth]{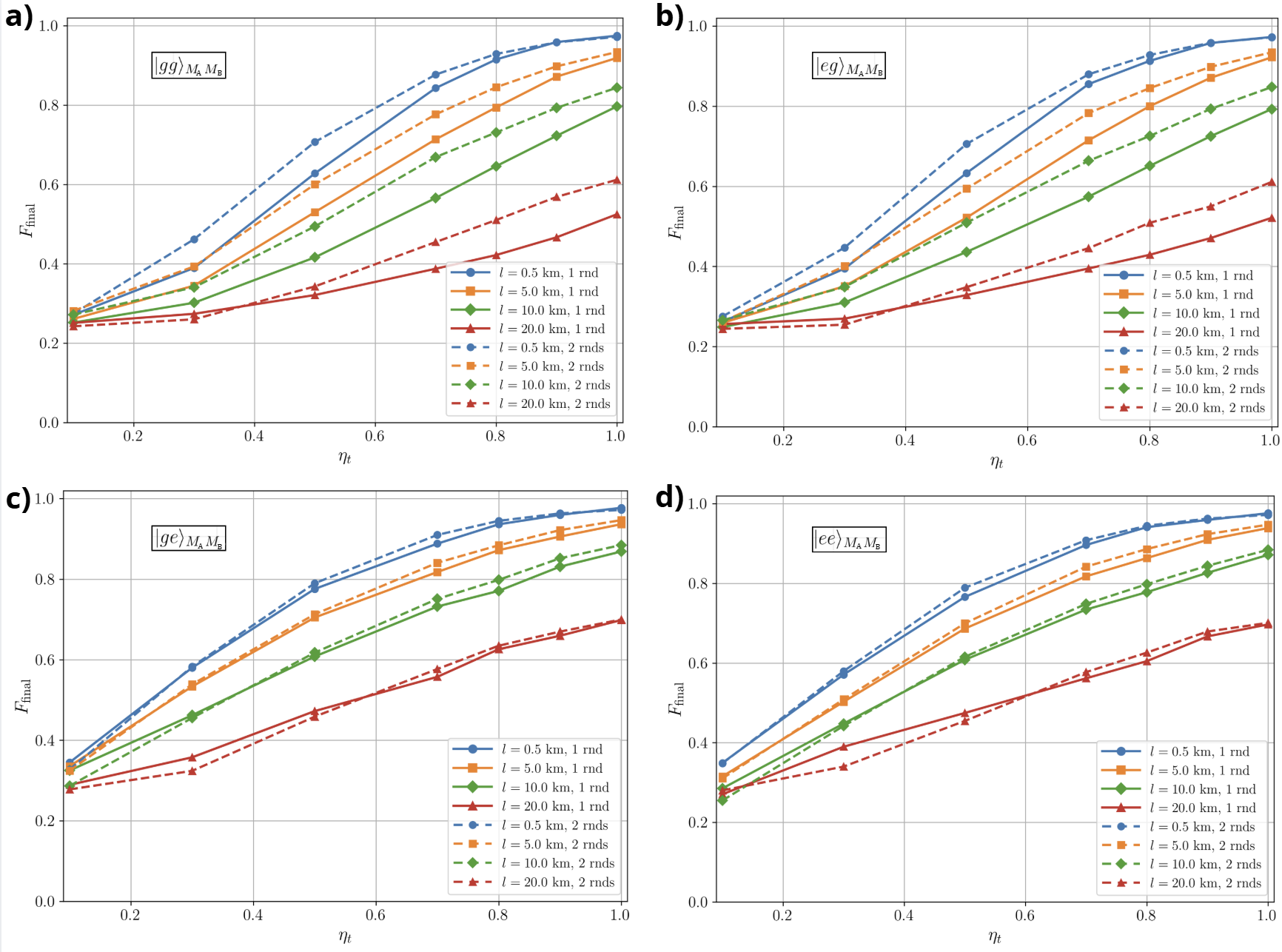}
    \caption{Fidelity from simulation results with single (\emph{1 rnd}) and double (\emph{2 rnds}) round of purification, shown as a function of transduction efficiency $\eta_\mathrm{t}$ for different fiber lengths. Each plot corresponds to a specific measurement outcome of $M_AM_B$: (a) $\ket{gg}_{M_AM_B}$, (b) $\ket{eg}_{M_AM_B}$, (c) $\ket{ge}_{M_AM_B}$, and (d) $\ket{ee}_{M_AM_B}$. $p_\mathrm{S}$ is set to 0.0002, and the remaining network parameters are as in Table~\ref{tab:simulation_parameters}.
}
    \label{fig:fidelityvalpurification}
\end{figure*}
In Fig.~\ref{fig:fidelityvalpurification}, we report the simulation results for the final fidelity of $R_AR_B$ qubits, for each $M_AM_B$ measurement outcome cases. Results are plotted as a function of the transduction efficiency and for different fiber lengths. We have applied both a single round and two rounds of purification using the previously introduced protocol. In the latter case, this procedure requires four initial instances of $\rho_{RM}$ on each source--TESQR link. \mhl{Additional purification rounds would cause a corresponding exponential scaling of initial resource cost.} To limit statistical fluctuations, we have performed $10^4$ simulation runs for each purification rounds and have computed the average of the resulting fidelities. For a fair comparison with the earlier performance analysis, we employ the same noise parameters (see in Table~\ref{tab:simulation_parameters}) and, analogously to the Hadamard-gate case, we fix the $S$ gate error probability at $p_\mathrm{S}=0.0002$~\cite{engineeringguide}.

Figure \ref{fig:fidelityvalpurification} demonstrates the efficacy of single-round purification. Across all four measurement outcomes, we observe a marked fidelity increase relative to the unpurified baseline (Fig. \ref{fig:fidelityvsetat}). The symmetries between outcomes $\ket{gg}_{M_AM_B}$ and $\ket{eg}_{M_AM_B}$ (and similarly $\ket{ge}_{M_AM_B}$ and $\ket{ee}_{M_AM_B}$) are preserved, and in the low-noise limit ($\eta_\mathrm{t} \to 1$, $l \to 0$ km), the performance across all the outcome scenarios becomes nearly identical. Quantitatively, the purification improvement is significant: for $l=0.5$ km and $\eta_\mathrm{t}=0.7$ [Figs. \ref{fig:fidelityvalpurification}(a,b)], fidelity rises from $\approx 0.65$ to over $0.85$. Likewise, for $l=5$ km and $10$ km, maximum fidelities reach $\approx 0.92$ and $0.80$, respectively (compared to pre-purification values of $0.75$ and $0.60$). At long distances, such as $l=20$ km, the critical fidelity value of $0.5$ is achievable, assuming unit transduction efficiency.

\begin{table*}[htb]
\centering
\caption{Simulation measurement outcome probabilities of $M_AM_B$ qubits, for single- and double-round purification at $l=10\ \mathrm{km}$ and for different transduction efficiencies. Each entry shows the probability pairs $[p(\cdot),p(\cdot)]$ for the corresponding grouped outcome $M_AM_B$. The other network parameters are set according to Table \ref{tab:simulation_parameters}. }
\renewcommand{\arraystretch}{1.7}

\begin{tabular}{ccccc} 
\hline\hline
  \multicolumn{1}{c}{} &
  \multicolumn{2}{c}{Single round} &
  \multicolumn{2}{c}{Double round} \\[4pt]
  $\eta_t$ &
  \makebox[4.0cm][c]{$[\ket{gg}_{M_AM_B},\ket{eg}_{M_AM_B}]$} &
  \makebox[4.0cm][c]{$[\ket{ge}_{M_AM_B},\ket{ee}_{M_AM_B}]$} &
  \makebox[4.0cm][c]{$[\ket{gg}_{M_AM_B},\ket{eg}_{M_AM_B}]$} &
  \makebox[4.0cm][c]{$[\ket{ge}_{M_AM_B},\ket{ee}_{M_AM_B}]$} \\
\hline
  0.10 & \makebox[3.0cm][c]{$[0.0979,\;0.1044]$} & \makebox[3.0cm][c]{$[0.0118,\;0.0101]$}
       & \makebox[3.0cm][c]{$[0.0074,\;0.0073]$} & \makebox[3.0cm][c]{$[0.0007,\;0.0009]$} \\
  0.30 & \makebox[3.0cm][c]{$[0.1047,\;0.1064]$} & \makebox[3.0cm][c]{$[0.0273,\;0.0283]$}
       & \makebox[3.0cm][c]{$[0.0089,\;0.0086]$} & \makebox[3.0cm][c]{$[0.0028,\;0.0028]$} \\
  0.50 & \makebox[3.0cm][c]{$[0.1116,\;0.1107]$} & \makebox[3.0cm][c]{$[0.0466,\;0.0459]$}
       & \makebox[3.0cm][c]{$[0.0123,\;0.0124]$} & \makebox[3.0cm][c]{$[0.0069,\;0.0069]$} \\
  0.80 & \makebox[3.0cm][c]{$[0.1150,\;0.1141]$} & \makebox[3.0cm][c]{$[0.0815,\;0.0826]$}
       & \makebox[3.0cm][c]{$[0.0265,\;0.0264]$} & \makebox[3.0cm][c]{$[0.0216,\;0.0214]$} \\
  1.00 & \makebox[3.0cm][c]{$[0.1272,\;0.1248]$} & \makebox[3.0cm][c]{$[0.1097,\;0.1092]$}
       & \makebox[3.0cm][c]{$[0.0464,\;0.0459]$} & \makebox[3.0cm][c]{$[0.0431,\;0.0404]$} \\ 
\hline\hline
\end{tabular}

\label{tab:outcome_probs_10km}
\end{table*}

Similar improvements are shown in Figs. \ref{fig:fidelityvalpurification}(c) and \ref{fig:fidelityvalpurification}(d). For $l=0.5$ km, a fidelity of $0.9$ is reached at $\eta_\mathrm{t}=0.7$, whereas the unpurified case required $\eta_\mathrm{t}>0.9$. Similarly, for $l=10$ km, the maximum fidelity increases to $0.88$ (from $0.7$), and even the $l=20$ km case improves from $0.6$ to $0.7$.

Notably, the transduction efficiency threshold for effective purification shifts to higher $\eta_\mathrm{t}$ as fiber length increases. This occurs because purification requires an input fidelity above $0.5$; as $l$ grows, the post-coupling fidelity drops, eventually falling below this critical value. We emphasize that this threshold cannot be inferred directly from Fig. \ref{fig:fidelityvsetat}, which depicts the final fidelity after swapping rather than the intermediate state immediately after coupling.

Regarding the double-round purification, the results in Fig. \ref{fig:fidelityvalpurification} show that it generally improves performance in all scenarios, with a more pronounced advantage when the process is dominated by noise, such as  in the measurement outcome cases $\ket{gg}_{M_AM_B}$ and $\ket{eg}_{M_AM_B}$, 
as previously discussed. However, there exist parameter regimes where double purification fails to outperform single-round purification and may even degrade performance. This effect occurs because the purification protocol enhances fidelity only for states with an initial fidelity above 0.5. If the output of the first round remains below this value, applying a second round—subject to realistic gate noise and decoherence—necessarily deteriorates the state.

Finally, since the purification protocol is probabilistic, it is interesting to examine the outcome probabilities associated with the four measurement outcomes of $M_AM_B$ qubits, both with one-round and two-round purification. As an illustrative example, Table~\ref{tab:outcome_probs_10km} reports these values as a function of the transduction efficiency for the case of fiber length $l=10$ km.

The numerical results reported in Table~\ref{tab:outcome_probs_10km} first confirm, under non-ideal transduction efficiency conditions, the expected imbalance in the outcome probabilities of the measurement results $\ket{gg}_{M_AM_B},\ket{eg}_{M_AM_B}$ compared to the other two. Moreover, as $\eta_t$ increases, the probabilities of both classes of outcomes tend to rise. In this regard, a more pronounced improvement is observed for the $\ket{ge}_{M_AM_B}, \ket{ee}_{M_AM_B}$ cases, where the absolute increase between the minimum and maximum values of $\eta_\mathrm{t}$ amounts to nearly one order of magnitude with a single purification round.  
A similar behavior is observed for the double-purification case. Here, the gain in fidelity is obtained at the cost of a global reduction of approximately one order of magnitude in the outcome probabilities. Nevertheless, with an ideal unit transduction efficiency, the four outcomes tend to become nearly equiprobable in both schemes. Even in such noisy scenarios, the overall outcome probability reaches maxima of $47\%$ and $18\%$ for single- and double-round purification, respectively. This confirms the feasibility of the proposed enhancement approach.

\section{Conclusion}
\label{sec:conclusion}

In this work, we introduce TESQR, a hybrid quantum repeater architecture that integrates optical-microwave transduction with quantum processing based on superconducting qubits. A key innovation of the proposed approach is the complete elimination of optical BSMs. Instead, entanglement swapping operations are directly performed in superconducting circuits. This architecture combines the excellent transmission capabilities of optical photons in classical communication channels with the high-fidelity quantum processing provided by superconducting matter qubits.

To evaluate the feasibility of this approach, we have analytically developed a comprehensive model that describes fiber optic transmission, transduction, coupling with superconducting qubits, and quantum computation. Realistic noise models are used for each component, leading to final fidelity expressions dependent on various non-ideal network parameters. Analytical results of the entanglement-swapping process demonstrate that, \mhl{within suitable operating regimes,} the proposed scheme can generate final entangled pairs with fidelities exceeding 0.5 over a total end-to-end distance of 20~km. Furthermore, it achieves average and maximum improvements of, respectively, 63\% and 159\% in entanglement distribution success probability compared to the corresponding photonic interference-based swapping scheme. \mhl{Unlike the latter, the TESQR approach always produces a final output state at the remote nodes, rather than aborting on failed photon-interference events. This operationally deterministic feature is advantageous because it preserves a usable state for subsequent processing, even when the output is not yet entangled.}

We further investigate the enhancement of performance by adding an entanglement purification stage applied 
at the repeater node. To assess the effectiveness of this method,
we have implemented a full network simulation using the QuTiP framework. Numerical tests with realistic device parameters resulted in end-to-end fidelity gains in excess of 0.8 for distances up to 20~km, both with single and double purification rounds.

Although current experimental demonstrations of microwave-optical transduction have not yet reached the efficiency levels required for immediate large-scale deployment, rapid progress and increasing interest in the field suggest that substantial improvements are achievable in the near term~\cite{nat1,nat2}. Therefore, the TESQR architecture provides a novel paradigm and a realistic pathway toward high-rate hybrid quantum repeaters, enabling efficient entanglement distribution in future quantum internet implementations. Future work will explore optimized purification strategies and potential performance enhancements based on quantum error correction techniques.

\section*{Data Availability Statement}
The simulator code is publicly available at~\cite{simulatorTESQR}. The authors will provide the data upon request.

\begin{acknowledgments}

This manuscript has been authored by the Fermi Forward Discovery Group, LLC under Contract No. 89243024CSC000002 with the U.S. Department of Energy, Office of Science, Office of High Energy Physics. The U.S. Department of Energy, Office of Science, Early Career Research Program supports S.Z., C.W., and F. F. (partially). J. W. and A.C. are supported by the Fermi Forward Discovery Group LLC under Contract No. FWP-23-24 with the U.S. Department of Energy, Office of Science, Advanced Scientific Computing Research (ASCR) Program. D.K. is supported by the U.S. Department of Energy, Office of Science, National Quantum Information Science Research Centers, Superconducting Quantum Materials and Systems Center (SQMS). The SQMS Center supports theory for quantum state preparation in superconducting quantum computing. 
 This work has been partially supported by the Italian Ministry of University and Research (MUR) in the framework of the FoReLab project (Departments of Excellence), that supported F.F., R.G., and M.P..

\end{acknowledgments}

\appendix
\section{Density-matrix expressions}
\label{app:densitymatrices}

The source joint state of the photonic mode \(T\) and the remote matter qubit \(R\) is:

\begin{equation}
|\psi\rangle_{TR} =
\frac{1}{\sqrt{2}}
\Bigl(
|0\rangle_T\otimes|g\rangle_R +
|1\rangle_T\otimes|e\rangle_R
\Bigr).
\end{equation}

Accordingly, the source state density matrix is:

\begin{equation}
\label{eq:rhoAB_source_simplified}
\begin{aligned}
\rho_{TR}^{\mathrm{source}}
&=
\frac{1}{2}
\Bigl(
|0\rangle_T\langle 0|_T\otimes|g\rangle_R\langle g|_R
+
|1\rangle_T\langle 1|_T\otimes|e\rangle_R\langle e|_R
\\
&\quad +
|0\rangle_T\langle 1|_T\otimes|g\rangle_R\langle e|_R
+
|1\rangle_T\langle 0|_T\otimes|e\rangle_R\langle g|_R
\Bigr).
\end{aligned}
\end{equation}

\subsection{Displaced-mode (reduced characteristic) operator}

The displacement operator on the photonic mode (of subsystem \(T\)) is \cite{olivares_introduction_2021}:

\[
D_T(\xi)=\exp\bigl(\xi a^\dagger - \xi^* a\bigr),
\]

and we use the matrix elements (in the \(|0\rangle,|1\rangle\) basis):

\begin{align}
X_{00}(\xi)&= \langle 0|D_T(\xi)|0\rangle = e^{-|\xi|^2/2},\\
X_{10}(\xi)&= \langle 1|D_T(\xi)|0\rangle = e^{-|\xi|^2/2}\,\xi,\\
X_{01}(\xi)&= \langle 0|D_T(\xi)|1\rangle = -e^{-|\xi|^2/2}\,\xi^*,\\
X_{11}(\xi)&= \langle 1|D_T(\xi)|1\rangle = e^{-|\xi|^2/2}\,(1-|\xi|^2).
\end{align}

We define the reduced characteristic operator on the qubit \(R\) by:

\begin{equation}
\rho_R^{\mathrm{source}}(\xi)=
\operatorname{Tr}_T
\!\left\{
\rho_{TR}^{\mathrm{source}}D_T(\xi)
\right\}.
\end{equation}

Computing the contributions term by term (using the source-state form \eqref{eq:rhoAB_source_simplified}), we get:

\begin{equation}
\label{eq:rhoB_source_xi}
\begin{split}
\rho_R^{\mathrm{source}}(\xi)
=
\frac{1}{2}e^{-|\xi|^2/2}
\Bigl(
|g\rangle_R\langle g|_R
+
(1-|\xi|^2)|e\rangle_R\langle e|_R
\\
+
\xi |g\rangle_R\langle e|_R
-
\xi^* |e\rangle_R\langle g|_R
\Bigr).
\end{split}
\end{equation}

\subsection{Fiber channel}

A pure-loss channel with transmissivity \(\eta_\mathrm{c}\) acts on the displacement as:

\[
D_T(\xi)
\rightarrow
D_T(\sqrt{\eta_\mathrm{c}}\xi)
e^{-\frac{1-\eta_\mathrm{c}}{2}|\xi|^2},
\]

so that:

\[
X_{mn}(\xi)
\mapsto
X_{mn}(\sqrt{\eta_\mathrm{c}}\xi)
e^{-\frac{1-\eta_\mathrm{c}}{2}|\xi|^2}.
\]

Applying this to \eqref{eq:rhoB_source_xi} gives:

\begin{equation}
\label{eq:rhoB_ch_xi}
\begin{split}
\rho_R^{\mathrm{ch}}(\xi)
&=
\frac{1}{2}e^{-|\xi|^2/2}
\Bigl(
|g\rangle_R\langle g|_R
+
(1-\eta_\mathrm{c}|\xi|^2)|e\rangle_R\langle e|_R
\\
&\quad
+
\sqrt{\eta_\mathrm{c}}\xi |g\rangle_R\langle e|_R
-
\sqrt{\eta_\mathrm{c}}\xi^* |e\rangle_R\langle g|_R
\Bigr).
\end{split}
\end{equation}

\subsection{Transduction Gaussian channel}

The transduction channel is modeled by transmissivity \(\eta_\mathrm{t}\) and added noise:

\[
N_\mathrm{t}=(1-\eta_\mathrm{t})\left(\bar n+\tfrac{1}{2}\right),
\]

with action:

\[
D_T(\xi)\mapsto
D_T(\sqrt{\eta_\mathrm{t}}\xi)e^{-N_\mathrm{t}|\xi|^2}.
\]

Substituting \(\xi\to\sqrt{\eta_\mathrm{t}}\xi\) in \eqref{eq:rhoB_ch_xi} and adding the extra damping factor yields:

\begin{equation}
\begin{split}
\rho_R^{\mathrm{tr}}(\xi)
&=
\frac{1}{2}
e^{-(\frac{\eta_\mathrm{t}}{2}+N_\mathrm{t})|\xi|^2}
\Bigl(
|g\rangle_R\langle g|_R
\\
&+
(1-\eta_\mathrm{c}\eta_\mathrm{t}|\xi|^2)
|e\rangle_R\langle e|_R
\\
&+
\sqrt{\eta_\mathrm{c}\eta_\mathrm{t}}\xi
|g\rangle_R\langle e|_R
-
\sqrt{\eta_\mathrm{c}\eta_\mathrm{t}}\xi^*
|e\rangle_R\langle g|_R
\Bigr).
\end{split}
\end{equation}

\subsection{Full System Density Matrix}

First of all, we derive the full system density matrix by using the Glauber formula \cite{olivares_introduction_2021}:

\begin{equation}
\rho_{TR}^{\text{tr}} =
\frac{1}{\pi}
\int
\rho_R^{\text{tr}}(\xi)
D_T^\dagger(\xi)
\, d^2\xi
\end{equation}

which explicitly produces:

\begin{widetext}

\begin{equation}
\begin{aligned}
\rho_{TR}^{\mathrm{tr}} =
&\frac{1}{2A}
|0\rangle_T\langle0|_T\otimes|g\rangle_R\langle g|_R
\\
&+
\frac12\left(\frac1A-\frac1{A^2}\right)
|1\rangle_T\langle1|_T\otimes|g\rangle_R\langle g|_R
\\
&+
\frac12\left(\frac1A-\frac{\eta_c\eta_t}{A^2}\right)
|0\rangle_T\langle0|_T\otimes|e\rangle_R\langle e|_R
\\
&+
\frac12\left(
\frac1A
-\frac{1+\eta_c\eta_t}{A^2}
+\frac{2\eta_c\eta_t}{A^3}
\right)
|1\rangle_T\langle1|_T\otimes|e\rangle_R\langle e|_R
\\
&+
\frac12
\sqrt{\eta_c\eta_t}
\frac1{A^2}
\Bigl[
|0\rangle_T\langle1|_T\otimes|g\rangle_R\langle e|_R
+
|1\rangle_T\langle0|_T\otimes|e\rangle_R\langle g|_R
\Bigr],
\end{aligned}
\end{equation}

\end{widetext}

where \(A=\frac{1+\eta_t}{2}+N_t\).

A new superconducting qubit $M$ is initialized in its ground state:
\[
\rho_{MTR}^{\mathrm{tr}} = |g\rangle_M\langle g|_M\otimes\rho_{TR}^{\mathrm{tr}}.
\]   

\subsection{Coupling}
\label{app:coupling-formula}
The joint-Kraus operators incorporating photon and qubit efficiencies $\eta_{\rm ph}$ and $\eta_{\rm qb}$ are defined as:
\begin{align}
    K_0 &= |g\,0\rangle_{MT}\langle g\,0|_{MT}- i \sqrt{\eta_{\rm qb} \eta_{\rm ph}} |e\,0\rangle_{MT}\langle g\,1|_{MT}, \\
    K_1 &= \sqrt{1 - \eta_{\rm qb} \eta_{\rm ph}} |g\,0\rangle_{MT}\langle g\,1|_{MT}.
\end{align}
The trace-preserving property can be readily verified as:
\begin{align}
    K_0^\dagger K_0 + K_1^\dagger K_1 = \mathbb{I}_{\rm sub},
\end{align}
where \(\mathbb{I}_{\rm sub}\) denotes the identity operator on the input subspace 
\(\{|g\,0\rangle, |g\,1\rangle\}\) (i.e., assuming that the superconducting qubit $M$ is prepared in the ground state).

The final state after applying these operators is given by:
\vspace{1em}
\begin{align}
    \rho_{MTR}^{\mathrm{coupl}} 
    = \sum_{i=0,1} \bigl(K_i \otimes \mathbb{I}_R \bigr) \,\rho_{MTR}^{\mathrm{tr}}\, \bigl(K_i^\dagger \otimes \mathbb{I}_R\bigr),
\end{align}
where $\mathbb{I}_R$ is the identity operator acting on subsystem $R$.

Collecting all transformed terms, the coupled output state is:
\begin{widetext}
\begin{equation}
\label{eq:rhocupled}
\begin{aligned}
    \rho_{MTR}^{\mathrm{coupl}} = &\frac{1}{2A} |g\,0\rangle_{MT}\langle g\,0|_{MT} \otimes |g\rangle_R\langle g|_R 
    + \frac{1}{2} \left( \frac{1}{A} - \frac{1}{A^2} \right) \Bigl[ \eta_{\rm qb} \eta_{\rm ph} |e\,0\rangle_{MT}\langle e\,0|_{MT} \otimes |g\rangle_R\langle g|_R \\
    &+ (1 - \eta_{\rm qb} \eta_{\rm ph}) |g\,0\rangle_{MT}\langle g\,0|_{MT} \otimes |g\rangle_R\langle g|_R \Bigr] 
    + \frac{1}{2} \left( \frac{1}{A} - \frac{\eta_\mathrm{c} \eta_\mathrm{t}}{A^2} \right) |g\,0\rangle_{MT}\langle g\,0|_{MT} \otimes |e\rangle_R\langle e|_R \\
    &+ \frac{1}{2} \left( \frac{1}{A} - \frac{1 + \eta_\mathrm{c} \eta_\mathrm{t}}{A^2} + \frac{2 \eta_\mathrm{c} \eta_\mathrm{t}}{A^3} \right) \Bigl[ \eta_{\rm qb} \eta_{\rm ph} |e\,0\rangle_{MT}\langle e\,0|_{MT} \otimes |e\rangle_R\langle e|_R \\
    &+ (1 - \eta_{\rm qb} \eta_{\rm ph}) |g\,0\rangle_{MT}\langle g\,0|_{MT} \otimes |e\rangle_R\langle e|_R \Bigr] \\
    &+ i \frac{1}{2} \sqrt{\eta_\mathrm{c} \eta_\mathrm{t}} \frac{\sqrt{\eta_{\rm qb} \eta_{\rm ph}}}{A^2} \Bigl[ |g\,0\rangle_{MT}\langle e\,0|_{MT}\otimes |g\rangle_R\langle e|_R 
    - |e\,0\rangle_{MT}\langle g\,0|_{MT} \otimes |e\rangle_R\langle g|_R \Bigr].
\end{aligned}
\end{equation}
\end{widetext}
Rewriting Eq.~(\ref{eq:rhocupled}) as: 
\(
\rho_{MRT}^{\mathrm{coupl}} = \rho_{MR}^{\mathrm{coupl}} \otimes \ket{0}_T\bra{0}_T,
\)
one observes that the photonic subsystem \(T\) is disentangled from the remaining subsystems \(M\) and \(R\). It can therefore be traced out from the global state and neglected in the subsequent analysis.

\subsection{Swapping and Fidelity Evaluation}

Recalling the symmetry of the two segments Alice--TESQR and Bob--TESQR (Section~\ref{sec:model}), we can assert that:
\(\rho_{MR}^{\mathrm{coupl}} = \rho_{M_AR_A}^{\mathrm{coupl}} = \rho_{M_BR_B}^{\mathrm{coupl}}\),
and obtain the full system density matrix as:
\begin{equation}
\label{eq:rho0}
\rho_o = \rho_{M_AR_AM_BR_B}^{\mathrm{coupl}}
= \rho_{M_AR_A}^{\mathrm{coupl}} \otimes \rho_{M_BR_B}^{\mathrm{coupl}}. 
\end{equation}

The subsequent calculations involving the application of noisy CNOT and Hadamard gates, followed by the noisy POVM measurement operation (as described in Section \ref{sec:swapping}), as well as the fidelity evaluation with respect to the corresponding ideal states, are not analytically tractable by hand. Instead, they are symbolically solved with the aid of the Python \texttt{Sympy} library \cite{meurer2017sympy}. These derivations are omitted here so as not to compromise readability.

\section{Thermal Noise in Electro-Optic Quantum Transduction}
\label{app:nbar}

\begin{figure}[t]
    \centering
    \includegraphics[width=\linewidth]{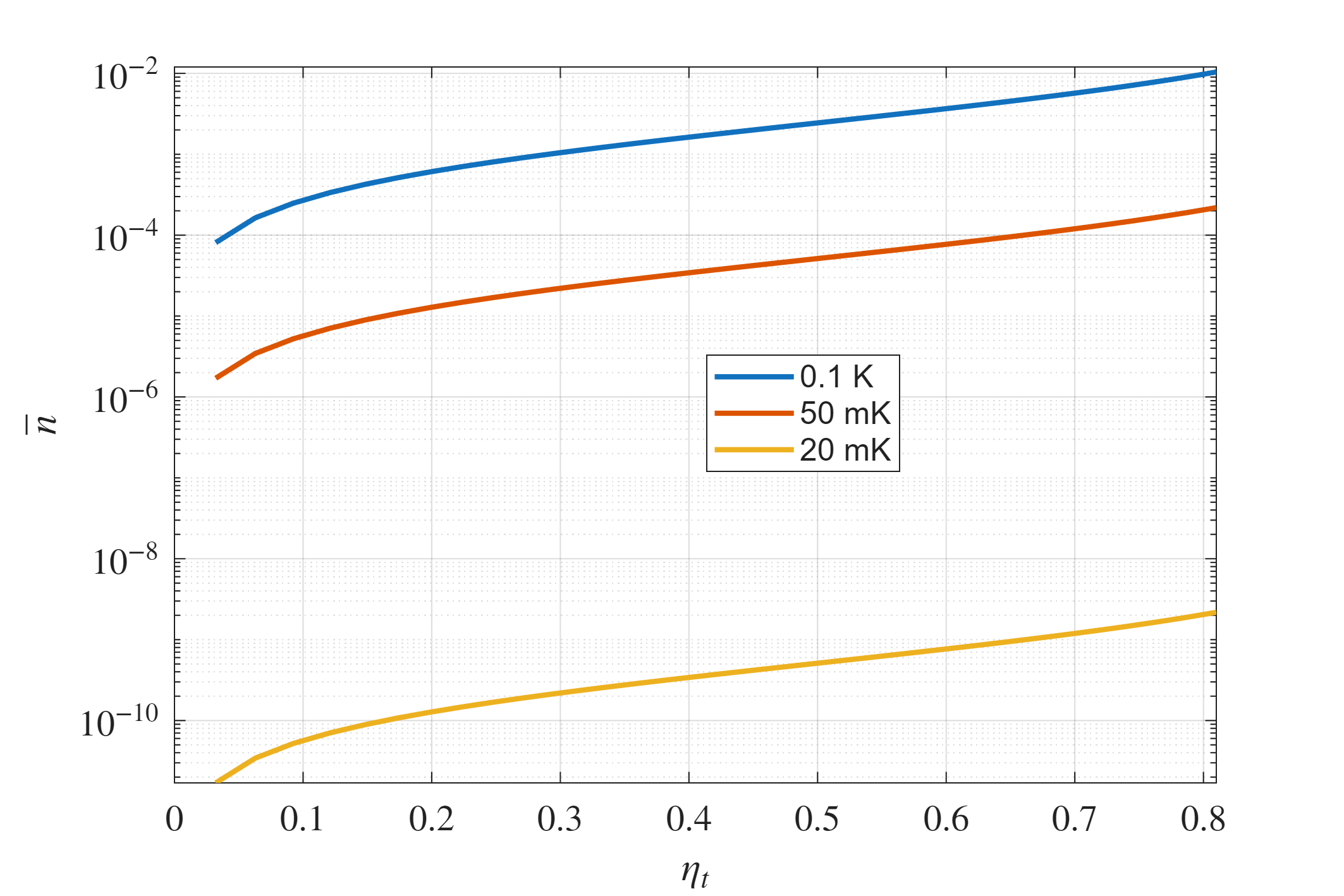}
    \caption{\mhl{Transducer thermal noise photon number $\bar{n}$ (logarithmic scale) as a function of the transduction efficiency $\eta_\mathrm{t}$ for different system temperatures $T_{\mathrm{sys}}$. The microwave and optical extraction efficiencies are fixed to $\zeta_\mathrm{m} = \zeta_\mathrm{o} = 0.9$, and a microwave frequency of $8~\mathrm{GHz}$ is assumed in the transducer.}}
    \label{fig:nbar}
\end{figure}

To assess realistic values of the thermal noise photon number $\bar{n}$ used throughout this work, we assume a fixed microwave extraction efficiency $\zeta_\mathrm{m} = 0.9$ and optical extraction efficiency $\zeta_\mathrm{o} = 0.9$, consistent with experimentally realistic parameters\mhl{~\cite{Rueda2019}}. Under these assumptions, Fig.~\ref{fig:nbar} shows the dependence of $\bar{n}$ on the transduction efficiency $\eta_\mathrm{t}$ for different system temperatures $T_{\mathrm{sys}}$, considering the electro-optic direct transduction scheme \mhl{without squeezing}~\cite{Jing}.

\mhl{We assume a representative microwave frequency of $8~\mathrm{GHz}$ when evaluating the Bose--Einstein thermal photon contribution in the transducer (with mean microwave occupation number $n_{\mathrm{in}}$).} The corresponding thermal occupation at optical frequencies is negligible~\cite{lauk2020perspectives}. Moreover, in the absence of squeezing, the added noise parameter $N_\mathrm{t}$ is invariant under the conversion direction, i.e., optical-to-microwave or microwave-to-optical~\cite{Jing2}.
\mhl{The overall transduction efficiency and output added noise parameter in such transduction scheme are expressed, respectively, by \cite{Jing2}:}
\mhl{
\begin{subequations}
    \begin{align}
        &\eta_{\mathrm{t}} = \zeta_{m} \zeta_{o} \frac{4C}{(1+C)^2}, \label{eq:eta_DC_electro}\\
    &N_{\mathrm{t}}=\frac{1}{2}+\frac{2C\zeta_o[2(1-\zeta_m)n_{\mathrm{in}}-\zeta_m]}{(1+C)^2}, \label{eq:N_DC_electro}
    \end{align} 
\end{subequations}
where $C$ indicates the interaction cooperativity ($\in [0,1)$).
Recalling the beam-splitter model expression \(N_{\mathrm{t}}=(1-\eta_{\mathrm{t}})\left(\bar{n}+\frac{1}{2}\right)\), it becomes evident from Equations (\ref{eq:eta_DC_electro}), (\ref{eq:N_DC_electro}) that the transduction efficiency and the mean thermal noise photon number are intrinsically linked.}

The absolute magnitude of $\bar{n}$ displayed in Fig. \ref{fig:nbar} is strongly temperature dependent. At $T_{\mathrm{sys}} = 0.1~\mathrm{K}$, \mhl{the thermal occupation reaches maximum values on the order of $10^{-2}$ at high $\eta_\mathrm{t}$}, whereas reducing the temperature to $50~\mathrm{mK}$ suppresses $\bar{n}$ by approximately two orders of magnitude over the entire efficiency range. At $T_{\mathrm{sys}} = 20~\mathrm{mK}$, the thermal contribution becomes negligible, with $\bar{n}$ remaining below $10^{-8}$ even at the largest transduction efficiencies considered.

\section{Local \(S\) Correction for Entanglement Purification}
\label{app:Spurifica}
The BBPSSW entanglement purification protocol \cite{Bennet} assumes as input a pair of two-qubit states of the Werner form. For two qubits the Werner state can be written as:
\begin{equation}
\label{eq:werner}
\rho_W(F) \;=\; F \ket{\Phi^{+}}\!\bra{\Phi^{+}} \;+\; \frac{1-F}{3}\!\sum_{\beta\in\{\Phi^{-},\Psi^{+},\Psi^{-}\}} \ket{\beta}\!\bra{\beta},
\end{equation}
where \(0\le F\le1\) denotes the fidelity with respect to the Bell state \(\ket{\Phi^{+}}=(\ket{00}+\ket{11})/\sqrt{2}\), and \(\{\ket{\Phi^{-}},\ket{\Psi^{\pm}}\}\) are the remaining Bell basis states.

In our notation the (ideal, noiseless) $R_AM_A$ state produced by the coupling process is:
\begin{equation}
\label{eq:ideal_state}
\ket{\psi}_{R_AM_A}
=\frac{1}{\sqrt{2}}\left(\ket{g g}_{R_AM_A} - i \ket{e e}_{R_AM_A}\right),
\end{equation}
where we identify \(\ket{g}\equiv\ket{0}\) and \(\ket{e}\equiv\ket{1}\) for the computational basis.

The single-qubit phase (``\(S\)'' ) gate is defined by the matrix:
\begin{equation}
\label{eq:Smatrix}
S \;=\; \begin{pmatrix} 1 & 0 \\[4pt] 0 & i \end{pmatrix},
\qquad
S^\dagger \;=\; \begin{pmatrix} 1 & 0 \\[4pt] 0 & -i \end{pmatrix}.
\end{equation}

Applying \(S\) on the second qubit (here: \(M_A\)), we obtain:
\begin{align}
\left(\mathbb{I}\otimes S\right)\ket{\psi}_{R_AM_A}
&= \frac{1}{\sqrt{2}}\big(\ket{g g}_{R_AM_A} - i \cdot i \ket{e e}_{R_AM_A}\big)\nonumber\\
&= \frac{1}{\sqrt{2}}\big(\ket{g g}_{R_AM_A} + \ket{e e}_{R_AM_A}\big)\nonumber\\
&=\ket{\Phi^{+}}. \label{eq:afterS}
\end{align}
Thus a local application of \(S\) transforms the state in Eq.~\eqref{eq:ideal_state} into the \(\ket{\Phi^{+}}\) Bell state, which is the canonical target for BBPSSW-style purification.

After performing purification, the local phase introduced by the \(S\) operation can be undone by applying \(S^\dagger\) on the same qubit. Indeed,
\begin{align}
\left(\mathbb{I}\otimes S^\dagger\right)\ket{\Phi^{+}}_{R_AM_A}
&= \frac{1}{\sqrt{2}}\big(\ket{g g}_{R_AM_A} + \ket{e}(-i)\ket{e}_{R_AM_A}\big)\nonumber\\
&= \frac{1}{\sqrt{2}}\big(\ket{g g}_{R_AM_A} - i \ket{e e}_{R_AM_A}\big),
\end{align}
which recovers the original state in Eq.~\eqref{eq:ideal_state}. 

\section{Role of the Photonic Truncation Dimension in the Simulator}
\label{app:dim_photonic}

In our simulation framework, the photonic Hilbert space is truncated to a finite Fock basis of dimension $\texttt{dim\_{photonic}}$. This parameter determines the maximum number of photonic excitations that can be explicitly represented in each photonic mode. The truncation is a practical necessity, as it reduces both memory usage and CPU requirements. 

\begin{figure}[htb]
    \centering
    \includegraphics[width=\linewidth]{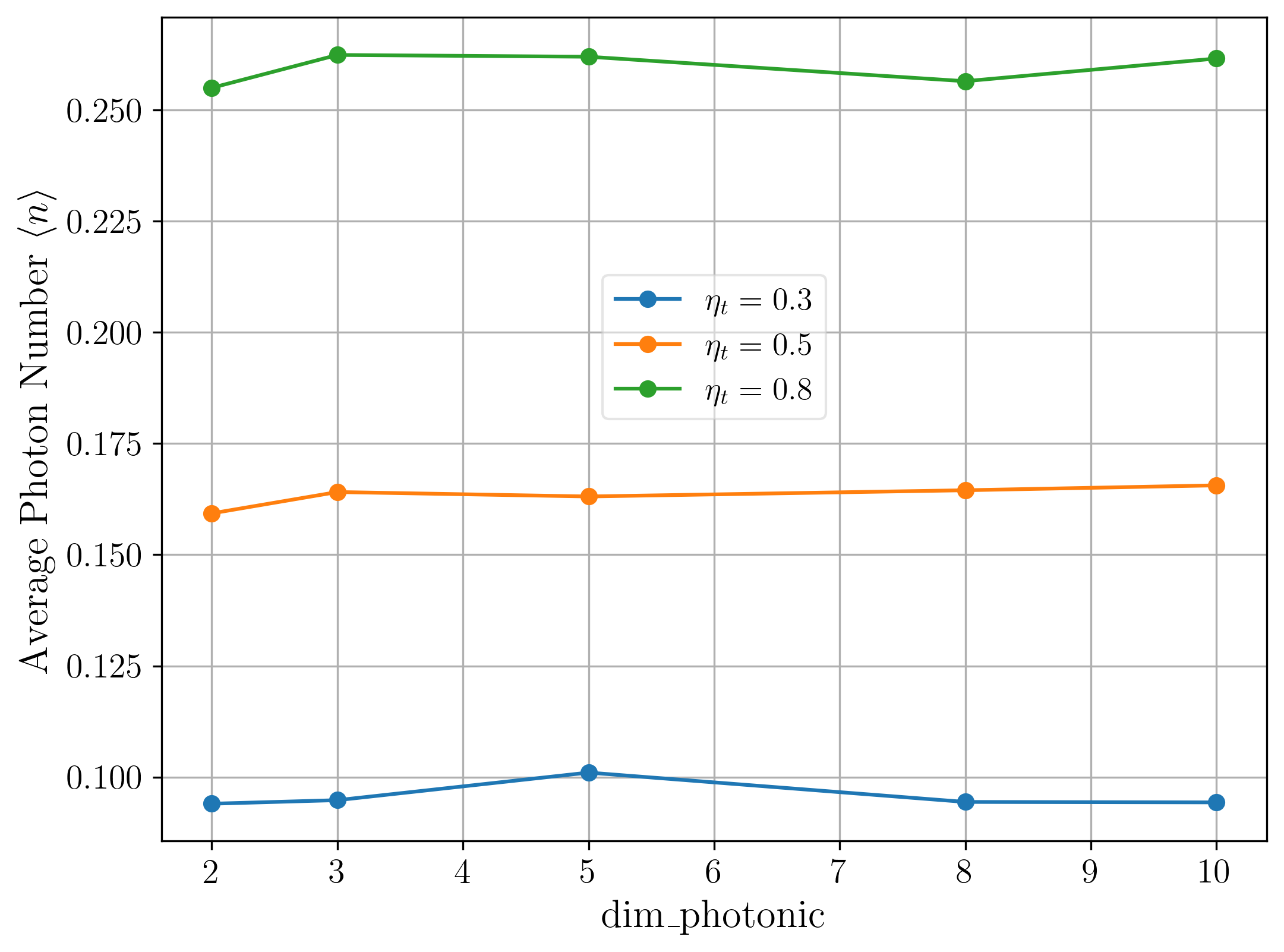}
    \caption{Average photon number $\langle n \rangle$ versus \texttt{dim\_photonic} after the transduction process, for various transduction efficiency values, with $l=10$ km and all other simulation parameters according to Table IV. }
    \label{fig:photon number}
\end{figure}

To quantify the effect of increasing the photonic truncation dimension, we have performed a dedicated test to verify variations in the photon number (in one of the two equivalent segments containing $T_A$ or $T_B$) after the transduction process. In Fig.~\ref{fig:photon number} we report the mean photon number $\langle n \rangle$ as a function of \texttt{dim\_photonic}, for different values of $\eta_\mathrm{t}$, with $l=10$ km and all other simulation parameters set according to Table~IV. To obtain averaged quantities, we have performed $10^5$ independent Monte Carlo simulation runs.
The numerical results in Fig.~\ref{fig:photon number} show that increasing \texttt{dim\_photonic} beyond two produces only a negligible change in the mean photon number after transduction: the increase remains smaller than 0.007, $0.008$, and 0.01 for, respectively, $\eta_\mathrm{t}=0.3$, $0.5$, and 0.8. Moreover, the trend is essentially stable, with maxima of 0.101, 0.164, and 0.262 for the three respective transduction efficiency values. Additional simulations further confirm that the empirical probability $p_{>1}$ (i.e., the probability that the photon number after transduction exceeds 1 in the simulation trials) remains approximately constant at $0.1\%$, $0.4\%$, and $1\%$ for $\eta_\mathrm{t}=0.3$, $0.5$, and $0.8$, respectively, when \texttt{dim\_photonic} is increased beyond 2. \mhl{Taken together, these results indicate that photon-addition events arising from the transduction model occur infrequently and are already well described within the \(\{|0\rangle,|1\rangle\}\) truncated basis for the considered operating regime. Moreover, this effect would be further enhanced by reducing the transduction system temperature \(T_{\mathrm{sys}}\).} From a practical standpoint, this justifies the use of $\texttt{dim\_photonic}=2$ as an efficient and reliable approximation, valid on average more than $99\%$ of the time in our network configuration.

\section{{Asymmetric Link Scenario}}
\label{app:asymmetric}

\mhl{
Throughout the main text, we have assumed a symmetric network configuration, in which the two segments Alice--TESQR and Bob--TESQR share identical physical parameters. This assumption was adopted to keep the analytical model tractable and to allow a clean assessment of the fundamental performance and feasibility of the TESQR scheme under standard operating conditions. It is also physically motivated: since all components of the TESQR architecture reside within the same integrated system, they are ideally engineered and controlled to exhibit matching properties, and deviations from symmetry can in principle be suppressed through careful calibration.}

\mhl{For completeness, we investigate here TESQR's entanglement-swapping fidelity performances when the two segments exhibit asymmetries in the following key physical parameters:
\begin{itemize}
    \item \textit{Fiber length asymmetry}: distinct lengths $l_A$ and $l_B$, yielding different channel transmissivities $\eta_{\mathrm{c},k} = e^{-l_k/l_{\mathrm{att}}}$, $k\in\{A,B\}$, where $k$ denotes Alice/Bob side;
    \item \textit{Transduction efficiency asymmetry}: distinct efficiencies $\eta_{\mathrm{t},A}$ and $\eta_{\mathrm{t},B}$, arising e.g.\ from fabrication tolerances or different cavity operating points;
    \item \textit{Thermal noise asymmetry}: distinct mean thermal photon numbers $\bar{n}_A$ and $\bar{n}_B$, reflecting different cryogenic bath temperatures $T_{\mathrm{sys},A}$ and $T_{\mathrm{sys},B}$.
\end{itemize}}
\mhl{All other parameters ($\eta_{\mathrm{qb}}$, $\eta_{\mathrm{ph}}$, $p_{\mathrm{CNOT}}$, $p_H$, $\varepsilon$) remain identical for the two segments, as they are performed locally on the same QPU.}

\subsection{\mhl{Model Extension}}
\mhl{
The derivation of each single-link density matrix proceeds as in Appendix~\ref{app:densitymatrices}, independently for each segment, with link-specific parameters $(\eta_{\mathrm{c},k},\,\eta_{\mathrm{t},k},\,\bar{n}_k)$. In particular, the transduction added-noise and normalization factor become:
\begin{equation}
    N_{\mathrm{t},k} = (1-\eta_{\mathrm{t},k})\!\left(\bar{n}_k+\tfrac{1}{2}\right), \qquad
    A_k = \frac{1+\eta_{\mathrm{t},k}}{2} + N_{\mathrm{t},k},
\end{equation}
for $k\in\{A,B\}$. All model steps up to and including the individual coupled density matrices remain formally identical to the symmetric case. The sole modification is in the assembly of the full four-partite state, where Eq.~\eqref{eq:rho0} is generalized to:
\begin{equation}
\label{eq:rho0_asym}
    \rho_o^{\mathrm{asym}} = \rho_{M_AR_A}^{\mathrm{coupl}} \otimes \rho_{M_BR_B}^{\mathrm{coupl}},
\end{equation}
with the two factors now in general distinct. The subsequent swapping circuit and fidelity evaluation proceed identically to the symmetric case, using \texttt{SymPy}~\cite{meurer2017sympy} for symbolic computation.}

\subsection{\mhl{Numerical Results and Discussion}}
\mhl{
We evaluate $F_{\mathrm{final}}$ after entanglement-swapping for three scenarios, each isolating one asymmetry parameter. The shared fixed parameters follow Table~IV, with central operating point $\eta_{\mathrm{t},0}=0.5$, $l_0=10~\mathrm{km}$, $T_{\mathrm{sys}}=0.1~\mathrm{K}$. Each asymmetry is introduced symmetrically as $x_A = x_0 - \Delta/2$, $x_B = x_0 + \Delta/2$, recovering the symmetric case at $\Delta=0$.}

\begin{figure}[htb]
    \centering
    \includegraphics[width=\linewidth]{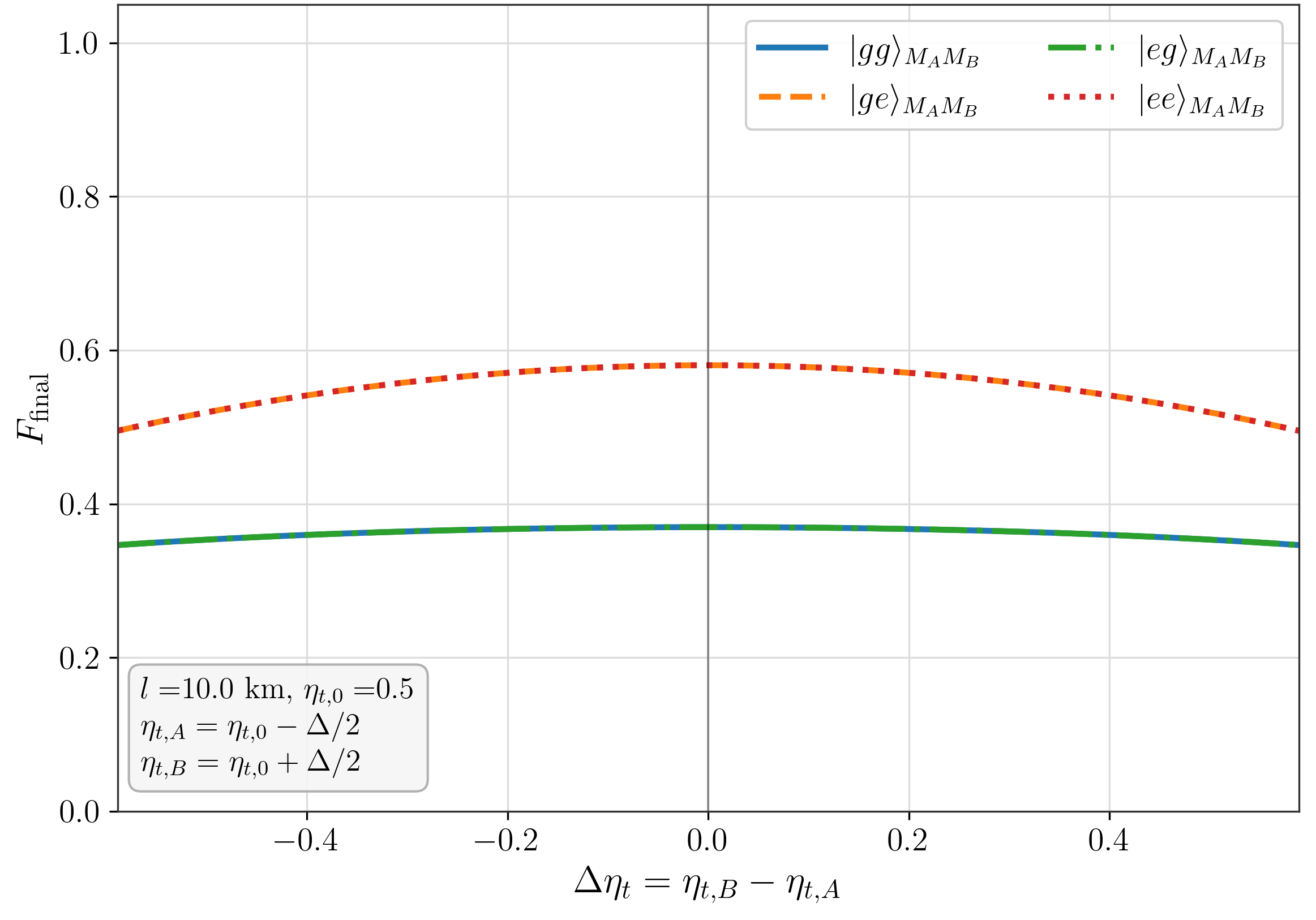}
    \caption{\mhl{Entanglement-swapping output fidelity ${F}_{\mathrm{final}}$ vs.\ transduction efficiency imbalance
    $\Delta\eta_t = \eta_{\mathrm{t},B}-\eta_{\mathrm{t},A}$,
    with $\eta_{\mathrm{t},A}=\eta_{\mathrm{t},0}-\Delta\eta_t/2$,
    $\eta_{\mathrm{t},B}=\eta_{\mathrm{t},0}+\Delta\eta_t/2$.
    Fixed: $l_0=10~\mathrm{km}$, $\eta_{\mathrm{t},0}=0.5$; other parameters as in Table~IV.}}
    \label{fig:asym_etat}
\end{figure}

\mhl{\textit{Transduction efficiency asymmetry.} Figure~\ref{fig:asym_etat} shows ${F}_{\mathrm{final}}$ versus $\Delta\eta_t = \eta_{\mathrm{t},B}-\eta_{\mathrm{t},A}$, with $\bar{n}_k$ derived self-consistently from $\eta_{\mathrm{t},k}$ via the transducer model (Appendix~\ref{app:nbar}).}

\mhl{All four outcomes degrade monotonically and symmetrically as $|\Delta\eta_t|$ grows. To understand this, we recall from Eq.~\eqref{eq:rhocupled} that the coupled density matrix of each link contains two structurally distinct contributions: diagonal population terms (proportional to $1/A_k$, $1/A_k^2$, $1/A_k^3$), which depend on $\eta_{\mathrm{t},k}$ through $A_k$; and an off-diagonal coherence term with coefficient:
\begin{equation}
    a_{5,k} = \frac{i}{2}\frac{\sqrt{\eta_{\mathrm{c},k}\,\eta_{\mathrm{t},k}}\,\sqrt{\eta_{\mathrm{qb}}\eta_{\mathrm{ph}}}}{A_k^2},
\end{equation}
which carries the quantum correlations between the matter qubit $M_k$ and the remote qubit $R_k$. Since $\eta_{\mathrm{t},k}$ appears in \textit{both} the diagonal and off-diagonal elements, any change in $\eta_{\mathrm{t},k}$ simultaneously affects all density-matrix entries for that link. The joint coherence contributing to the final entangled state after swapping is therefore proportional to $\sqrt{\eta_{\mathrm{t},A}\cdot\eta_{\mathrm{t},B}}$. Since $\eta_{\mathrm{t},A}+\eta_{\mathrm{t},B}=2\eta_{\mathrm{t},0}$ is fixed, the AM--GM inequality gives:
\begin{equation}
    \sqrt{\eta_{\mathrm{t},A}\cdot\eta_{\mathrm{t},B}} \;\leq\; \eta_{\mathrm{t},0},
\end{equation}
with equality only when $\eta_{\mathrm{t},A}=\eta_{\mathrm{t},B}$. Because $\eta_{\mathrm{t}}$ affects all density-matrix elements, this suppression acts uniformly on all four measurement outcomes, explaining the identical qualitative behavior seen in Fig.~\ref{fig:asym_etat}.}

\begin{figure}[htb]
    \centering
    \includegraphics[width=\linewidth]{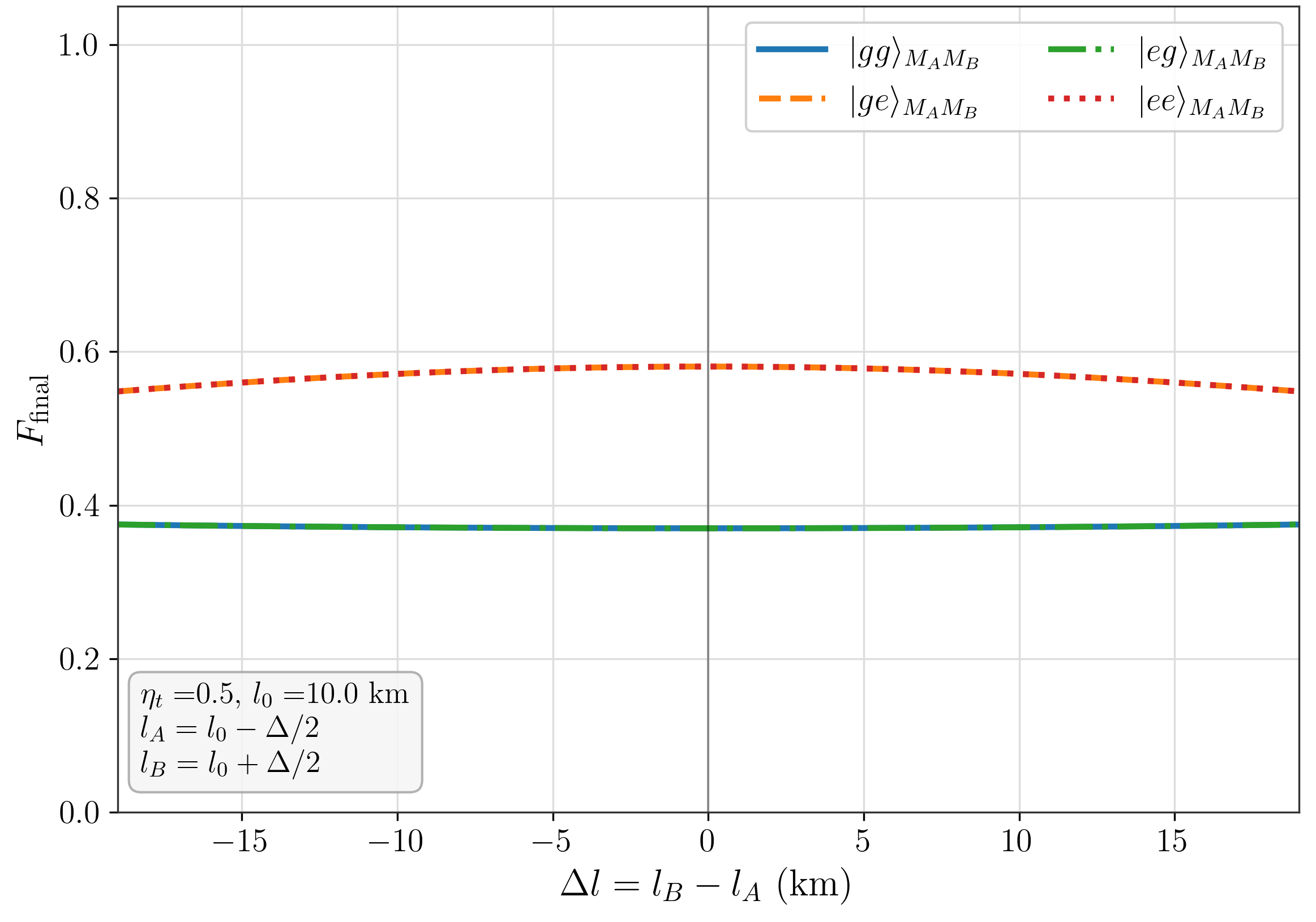}
    \caption{\mhl{Entanglement-swapping output fidelity ${F}_{\mathrm{final}}$ vs.\ fiber length imbalance $\Delta l = l_B-l_A$,
    with $l_A=l_0-\Delta l/2$, $l_B=l_0+\Delta l/2$.
    Fixed: $\eta_{\mathrm{t},0}=0.5$, $l_0=10~\mathrm{km}$; other parameters as in Table~IV.}}
    \label{fig:asym_l}
\end{figure}

\mhl{\textit{Fiber length asymmetry.} Figure~\ref{fig:asym_l} shows ${F}_{\mathrm{final}}$ versus $\Delta l = l_B - l_A$, with $\eta_{\mathrm{t},A}=\eta_{\mathrm{t},B}=\eta_{\mathrm{t},0}$ fixed.}

\mhl{
Unlike the $\Delta\eta_t$ case, this scenario reveals a \textit{split} behavior: outcomes 
$|ge\rangle_{M_AM_B}$ and $|ee\rangle_{M_AM_B}$ degrade with increasing $|\Delta l|$, while 
$|gg\rangle_{M_AM_B}$ and $|eg\rangle_{M_AM_B}$ remain nearly flat. To understand this, 
recall that the target states for the latter two outcomes are $|gg\rangle\pm|ee\rangle$, 
whose fidelity depends on the probability of both remote qubits $R_A$, $R_B$ being found in 
the same state ($|g\rangle$ or $|e\rangle$). This is governed by the diagonal coefficients 
$a_{1,k}$ and $a_{2,k}$ of Eq.~\eqref{eq:rhocupled}, which are independent of 
$\eta_{\mathrm{c},k}$ since the normalization factor $A_k$ depends only on $\eta_{\mathrm{t},k}$ 
and $\bar{n}_k$. These terms correspond to events in which the photon is lost before 
reaching $R_k$, so their magnitude is set entirely by the transducer and is unaffected 
by fiber attenuation. As a consequence, varying $\Delta l$---which only modifies 
$\eta_{\mathrm{c},k}$ while leaving $A_k$ unchanged---does not alter $a_{1,k}$ or $a_{2,k}$, 
and the fidelity of $|gg\rangle$ and $|eg\rangle$ remains flat. The target states 
$|ge\rangle\pm|eg\rangle$ for the other two outcomes instead require quantum coherence 
between $|g\rangle_R$ and $|e\rangle_R$ on each link, carried by the off-diagonal term 
$a_{5,k}\propto\sqrt{\eta_{\mathrm{c},k}\eta_{\mathrm{t},k}}/A_k^2$ and by the 
$\eta_{\mathrm{c},k}$-dependent diagonal terms $a_{3,k}$ and $a_{4,k}$. These coefficients 
are directly affected by fiber asymmetry, and their joint degradation across the two links 
causes the fidelity of $|ge\rangle$ and $|ee\rangle$ to decrease with $|\Delta l|$.
This behavior contrasts sharply with the $\Delta\eta_t$ case: there, $\eta_{\mathrm{t},k}$ 
enters $A_k$ directly, so any transducer imbalance simultaneously modifies all 
coefficients $a_{1,k}$ through $a_{5,k}$, degrading all four outcomes uniformly.}

\begin{figure}[htb]
    \centering
    \includegraphics[width=\linewidth]{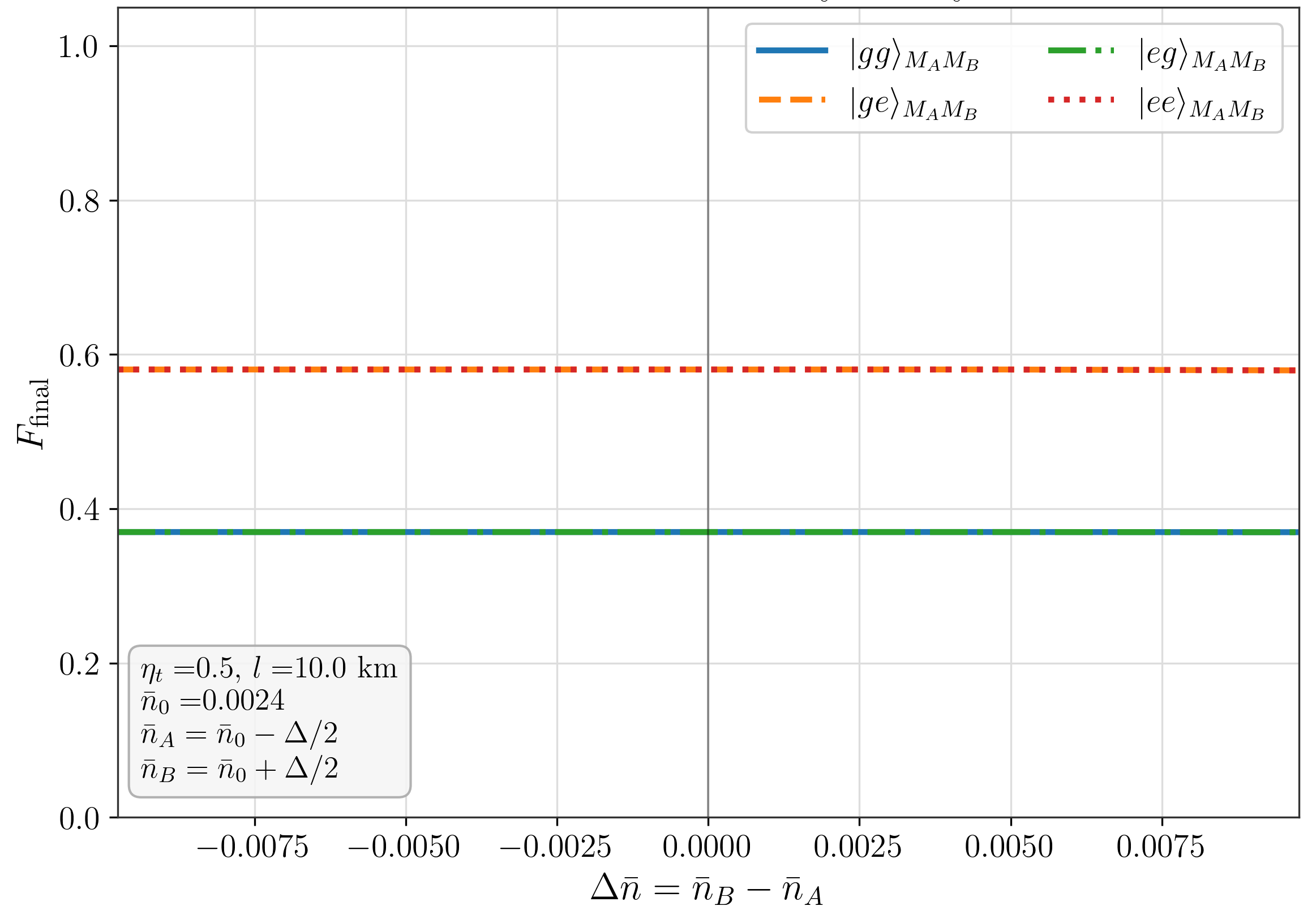}
    \caption{\mhl{Entanglement-swapping output fidelity ${F}_{\mathrm{final}}$ vs.\ thermal noise imbalance $\Delta\bar{n}=\bar{n}_B-\bar{n}_A$,
    with $\bar{n}_A=\bar{n}_0-\Delta\bar{n}/2$, $\bar{n}_B=\bar{n}_0+\Delta\bar{n}/2$.
    Fixed: $\eta_{\mathrm{t},0}=0.5$, $l_0=10~\mathrm{km}$; other parameters as in Table~IV.}}
    \label{fig:asym_nbar}
\end{figure}

\mhl{\textit{Thermal noise asymmetry.} Figure~\ref{fig:asym_nbar} shows ${F}_{\mathrm{final}}$ versus $\Delta\bar{n}=\bar{n}_B-\bar{n}_A$, with $\eta_{\mathrm{t}}$ and $l$ fixed at their nominal values.}

\mhl{
All four curves are essentially flat. At $T_{\mathrm{sys}}=0.1~\mathrm{K}$ and $f=8~\mathrm{GHz}$, the thermal occupation $\bar{n}_0 \approx 2.4\times10^{-3}$ is negligible compared to the vacuum term in $N_{\mathrm{t}}=(1-\eta_{\mathrm{t}})(\bar{n}+1/2)$. The absolute perturbation on $A_k$ induced by the full sweep of $\Delta\bar{n}$ is of order $(1-\eta_{\mathrm{t}})\bar{n}_0\sim 10^{-3}$, entirely masked by the contributions from $\eta_{\mathrm{t}}$, $\eta_{\mathrm{c}}$, and the gate errors. This confirms that, at the cryogenic temperatures of interest, thermal noise is not a limiting factor of the TESQR scheme, and moderate inter-segment temperature differences have no measurable impact on fidelity. The dominant sources of degradation remain $\eta_{\mathrm{t}}$, $\eta_{\mathrm{c}}$, and gate imperfections, consistently with the main text analysis. These results jointly support the physical validity and robustness of the symmetric scenario adopted as the primary framework throughout this work. }



\bibliography{biblio}


\end{document}